\documentclass[12pt, draftclsnofoot, onecolumn]{IEEEtran}
\usepackage{algorithm,amsbsy,amsmath,amssymb,epsfig,bbm,mathrsfs,multirow,amsthm}
\usepackage{color}
\usepackage{algpseudocode,tabularx}
\usepackage{graphicx,caption,subcaption,array,multirow}
\usepackage{makecell}
\usepackage[hidelinks]{hyperref}

\renewcommand{\baselinestretch}{1.355}

\DeclareMathAlphabet{\mathpzc}{OT1}{pzc}{m}{it}
\DeclareMathOperator*{\argmax}{\arg\!\max}

\newtheorem{theorem}{\textbf{\textsc{Theorem}}}

\makeatletter
\newcommand{\multiline}[1]{%
	\begin{tabularx}{\dimexpr\linewidth-\ALG@thistlm}[t]{@{}X@{}}
		#1
	\end{tabularx}
}
\makeatother
\begin{document}
	
\title{DeepFake: Deep Dueling-based Deception Strategy to Defeat Reactive Jammers}
\author{Nguyen Van Huynh, Dinh Thai Hoang, Diep N. Nguyen, and Eryk Dutkiewicz
	\thanks{Nguyen Van Huynh, Dinh Thai Hoang, Diep N. Nguyen, and Eryk Dutkiewicz are with University of Technology Sydney, Australia. E-mails: huynh.nguyenvan@student.uts.edu.au, \{Hoang.Dinh, Diep.Nguyen, and Eryk.Dutkiewicz\}@uts.edu.au.}
}
\maketitle
\thispagestyle{empty}

\begin{abstract}
In this paper, we introduce DeepFake, a novel deep reinforcement learning-based deception strategy to deal with reactive jamming attacks. In particular, for a smart and reactive jamming attack, the jammer is able to sense the channel and attack the channel if it detects communications from the legitimate transmitter. To deal with such attacks, we propose an intelligent deception strategy which allows the legitimate transmitter to transmit ``fake'' signals to attract the jammer. Then, if the jammer attacks the channel, the transmitter can leverage the strong jamming signals to transmit data by using ambient backscatter communication technology or harvest energy from the strong jamming signals for future use. By doing so, we can not only undermine the attack ability of the jammer, but also utilize jamming signals to improve the system performance. To effectively learn from and adapt to the dynamic and uncertainty of jamming attacks, we develop a novel deep reinforcement learning algorithm using the deep dueling neural network architecture to obtain the optimal policy with thousand times faster than those of the conventional reinforcement algorithms. Extensive simulation results reveal that our proposed DeepFake framework is superior to other anti-jamming strategies in terms of throughput, packet loss, and learning rate. 
\end{abstract}



\section{Introduction}
Due to the open and broadcast nature of wireless links, wireless communications are extremely vulnerable to jamming attacks, especially for low-power systems such as Internet of Things (IoT). In practice, the jammer can easily launch attacks by injecting high-power interference to the target communication channel~\cite{Hanawal2016Joint}. Consequently, the signal-to-interference-plus-noise ratio (SINR) at the receiver can be significantly reduced, and thus the receiver may not be able to decode the information sent from the transmitter. Among radio jamming methods, dealing with reactive jamming is very challenging as the jammer can ``smartly'' attack the channel whenever it detects transmissions from the transmitter on the channel. In addition, detecting reactive jammer is more difficult as the detector might not be able to distinguish between the jamming signals and the signals sent from the transmitter. More importantly, reactive jamming can be easily launched by conventional jammers by equipping off-the-shelf signal-detection circuits. Thus, reactive jamming attacks can cause serious consequences in critical communications systems such as military, medical, and public safety. As such, defeating reactive jamming attacks has been an urgent mission for future wireless communication networks.
\subsection{Current Solutions and Limitations}
Various anti-jamming solutions have been proposed in the literature. Nevertheless, these solutions are not effective in dealing with reactive jammers. In this section, we will study existing anti-jamming solutions together with their limitations in dealing with reactive jamming attacks.
\subsubsection{Regulating Transmit Power}
Regulating the transmit power at the legitimate transmitter is the simplest solution and was introduced from the early days of dealing with jamming attacks~\cite{Mpitziopoulos2009Survey}. In particular, a transmitter can choose to transmit at a very low power level so that the jammer cannot detect its transmission. However, in this way, the jammer can always force the transmitter to transmit data at a very low rate, and the transmitter even cannot transmit data if the jammer is equipped with a very sensitive signal-detection circuit. Another solution for the transmitter is transmitting signals at a very high power level to dominate jamming signals. Nevertheless, this solution possesses several drawbacks. First, increasing the transmit power introduces a new problem as the transmitter can cause unintentional interference to other nearby radio systems~\cite{Xu2006Jamming}. Second, with reactive jammers which can sense the legitimate transmissions and adjust its attack strategy, e.g., increase the attack power level, transmitting signals at high power levels is not an effective way. Finally, if the jammer has a sufficient power budget, it can always disrupt all the ongoing transmissions.
\subsubsection{Frequency Hopping}
Frequency-hopping spread spectrum (FHSS) is a common technology when dealing with jamming attacks~\cite{Quan2015Anti}-\cite{Wu2011Anti}. The key idea of this technique is using a switching algorithm that allows the transmitter and receiver to find a new channel for communications once the current channel is attacked by the jammer. In~\cite{Mpitziopoulos2007Hybrid}, the authors introduced a novel hybrid mechanism to cope with fast-following jammers by combining FHSS and direct sequence spread spectrum (DSSS) technologies. In particular, each frequency channel implements the DSSS modulation with a 16-bit Pseudo noise code. Differently, the authors in~\cite{Wang2011Anti} aimed to avoid jamming attacks by introducing a stochastic game framework. Then, the minimax Q-learning algorithm is adopted to obtain the optimal defense policy for the legitimate transmitter (i.e., how to switch between channels). Similarly, game theory based anti-jamming frameworks with FHSS technology are also introduced in~\cite{Gao2018Game} and~\cite{Wu2011Anti}.

However, the above solutions and others in the literature possess several limitations in dealing with reactive jamming attacks. First, when the transmitter hops to a new channel, the reactive jammer also can discern/sense the transmitter's activities to attack the new channel. Second, the FHSS schemes require multiple available channels for communications at the same time and a predefined switching algorithm implemented on both the transmitter and receiver. As such, this solution may not be feasible to widely implement on resource-constrained and channel-limited wireless systems. More importantly, if the jammer has sufficient energy to attack all the channels simultaneously, the FHSS schemes do not work anymore. In addition, the game models proposed in~\cite{Wang2011Anti},~\cite{Gao2018Game}, and~\cite{Wu2011Anti} may not be effective as the environment information is required as the input of the algorithm. Unfortunately, in practice, it is difficult to obtain the environment information in advance, especially when dealing with the reactive jammer which can adjust its attack strategy by sensing the transmitter's transmissions.
\subsubsection{Rate Adaptation}
Another countermeasure to prevent and mitigate impacts of jamming attacks is the rate adaptation (RA) technique~\cite{Pelechrinis2009RA}-\cite{Noubir2011RA}. The RA technique allows the transmitter to reduce its transmission rate when the jammer attacks the channel. The reason is that under jamming attacks, the channel condition is not good with interference from the jammer. Thus, reducing the data rate is a potential solution as a lower rate is more reliable and suitable for poor channel quality~\cite{Noubir2011RA}. However, this technique possesses several limitations. The authors in~\cite{Firouzbakht2012RA} and~\cite{Noubir2011RA} demonstrated that the RA technology possesses low performance on a single channel and in dealing with reactive jamming attacks. In particular, this technique assumes that the transmitter can observe the actual jammer's attack performance before selecting an appropriate transmission rate at which the receiver can successfully decode the information. However, for reactive jammers, they only attack the channels after the transmitter transmits data, and thus the RA technique is not effective in dealing with the reactive jammers.
\subsubsection{Recent Solutions}
Recently, there are some new ideas introduced to deal with jamming attacks which are especially efficient for low-power systems.  Specifically, the authors in~\cite{Gu02017Exploiting} proposed the idea of harvesting energy from jamming signals. This is stemmed from the fact that the jammers usually use high transmission power levels to disturb legitimate communications, and thus the limited-energy devices (e.g., IoT devices) can harvest an abundant energy from the jammers by using RF energy harvesting techniques. In~\cite{Zhao2017Antijamming}, the authors proposed two schemes to defeat jamming attacks in interference alignment (IA) networks. In particular, in the first scheme, the interference and the jamming signal at the receiver are aligned into the same subspace, and thus they can be eliminated. Then, the second scheme is proposed to further enhance the anti-jamming performance by maximizing the SINR at the receiver. In~\cite{Guo2019Proactive}, the authors pointed out that the jamming signals can be beneficial for IA networks with the presence of eavesdroppers. In particular, a proactive jammer is designed to disrupt the eavesdropping while the receiver can avoid the jamming signals by proactively aligning the jamming signal into the same subspace as that of the interference. In~\cite{Huynh2019Jam}, the authors introduce a new approach of using ambient backscatter technology~\cite{Liu2013Ambient} to deal with jamming attacks. The key idea of this approach is that when the jammer attacks the channel, the transmitter can backscatter the jamming signals to transmit data to the receiver. Under this solution, the transmitter does not need to ``hide'' or ``escape'' from the jammer as it can leverage the strong jamming signals for its transmissions. Interestingly, the transmitter can harvest more energy and backscatter more data when the jammer attacks the channel with higher power levels. However, this solution only works well with proactive jammers because the transmitter only can harvest energy or backscatter from jamming signals after the jammers attack the channel. For reactive jammers, they will not attack if there is no activity of the transmitter on the channel, and thus this solution is not applicable to defeat the reactive jammers. Given the above, dealing with reactive jammers is very challenging, especially for low-power communication systems and when the jammer's power budget is high.

For that, in~\cite{Hoang2020Borrowing}, the authors propose a deception mechanism to defeat reactive jammers for IoT networks. In particular, the transmitter can decide either to perform deception or actively transmit data at the beginning of the time slot. If the transmitter chooses to perform deception, it will first generate ``fake'' signals to lure the jammer. After that, as soon as the jammer attacks the channel, the transmitter can harvest RF energy from the jamming signals or backscatter information to the receiver through the jamming signals. However, this solution adopts the conventional MDP framework with only one decision epoch per time slot. Consequently, this framework cannot effectively deal with reactive jamming attacks because decisions (e.g., harvest energy, backscatter and rate adaption) must be performed before the jammer attacks the channel. In contrast, our proposed two-decision epoch MDP framework can maximize the efficiency in defeating the reactive jammer as it can let the transmitter to choose the best actions by observing the actual status of the jammer after it attacks the channel. Moreover, the jammer strategy considered in~\cite{Hoang2020Borrowing} uses only one power level, and thus it is much easier for the transmitter to learn and find the optimal policy. To the best of our knowledge, all current anti-jamming approaches cannot efficiently deal with reactive jamming attacks.
\subsection{Main Contributions}
In this paper, we develop an intelligent anti-jamming framework to cope with a powerful reactive jammer which can attack the channel once it detects active transmissions from the transmitter. In particular, we first introduce a deception mechanism that enables the transmitter to lure the jammer by actively transmitting signals for a short period of time. After that, if the jammer attacks the channel, we propose the ideas that allow the transmitter to either harvest energy from the jamming signal, backscatter jamming signals to transmit data using ambient backscatter technology, or actively transmit data based on RA technique. Moreover, to deal with the dynamic and uncertainty of jamming attacks, we develop a new Markov decision process framework with two decision epochs over one time slot to formulate the anti-jamming deception strategy for the transmitter and then use the Q-learning algorithm to obtain the optimal defense policy for the transmitter. Although the Q-learning algorithm is an effective tool to help the transmitter obtain the optimal policy without requiring jammer's information in advance, its convergence is usually very slow and might not be efficient to implement in practice. Thus, we develop a novel deep dueling reinforcement learning algorithm that enables the transmitter to obtain the optimal policy thousand times faster than those of the conventional reinforcement learning methods, e.g., Q-learning and deep Q-learning algorithms. The key idea of this algorithm is to separately estimate the advantage and value functions of each state-action pair. In this way, the learning rate can be significantly improved. It is worth noting that the reactive jammer can adjust its attack policy by sensing the activities of the transmitter on the channel, e.g., actively transmit or idle. Thus, with a very fast convergence rate, our proposed solution can quickly and efficiently adapt the optimal defense strategy when the jammer adjusts its policy. Extensive simulation results then show that our proposed solution can achieve a very good performance in terms of throughput and packet loss compared with other conventional anti-jamming strategies. Interestingly, we show that with our proposed solution, the transmitter can utilize the power from jamming signals, and thus the more power the jammer uses to attack the channel, the greater performance we can achieve. The key contributions of the paper are summarized as follows:
\begin{itemize}
	\item Propose an intelligent anti-jamming deception strategy that can undermine the jammer's attack ability and leverage the jammer's power to enhance the system performance.
	\item Introduce novel ideas of using RF energy harvesting and ambient backscatter techniques which can further exploit jamming signals to achieve greater performance.
	\item Develop a new dynamic MDP model to deal with reactive jamming attacks and propose the reinforcement algorithm to help the transmitter obtain the optimal defense policy without requiring information about jammer in advance.
	\item Develop a deep reinforcement learning algorithm with a new neural network architecture which can capture the properties of our considered system and quickly find the optimal defense policy for the transmitter.
	\item Perform extensive simulations to show the efficiency of our proposed solutions as well as to study key factors which have significant impacts in defeating reactive jamming attacks.
\end{itemize}
The rest of this paper is organized as follows. In Section~\ref{Sec.System}, we describe the anti-jamming system model. Then, the formulation of the proposed dynamic MDP is presented in Section~\ref{Sec:prob}. In Section~\ref{sec:QDeepQ}, we introduce the Q-learning and deep Q-learning algorithms. Then, the deep dueling Q-learning algorithm is presented in Section~\ref{Sec:deepdueling}. Section~\ref{sec:evaluation} presents the simulation results. Finally, Section~\ref{sec:conclusion} concludes the paper.
\section{System Model}
\label{Sec.System}
\begin{figure*}[!]
	\centering
	\includegraphics[scale=0.08]{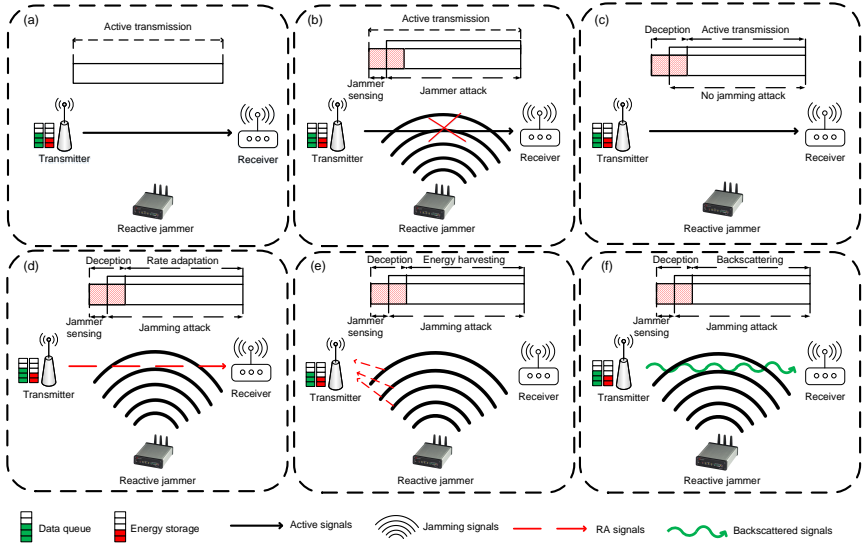}
	\caption{An illustration of deception strategy to deal with reactive jamming attacks.}
	\label{Fig.system_model}
\end{figure*}
This paper considers a wireless communication system in which a legitimate transmitter communicates with a receiver through a dedicated channel $C$ as shown in Fig.~\ref{Fig.system_model}. In particular, the transmitter is equipped an RF energy harvesting circuit to harvest energy from surrounding signals, e.g., jamming signals. The harvested energy is then stored in an energy storage with the maximum size of $E$ energy units. We assume that time is slotted. At each time slot, we assume that the transmitter can harvest $e_v$ units of energy from surrounding RF signals with probability $p_e$. The transmitter is also equipped with a data queue. The maximum data queue size is denoted by $D$. When a new packet arrives at the transmitter, if the data queue is not full, the packet will be stored in the queue. Otherwise, the packet will be dropped. At each time slot there are $K$ packets arriving at the data queue with probability $\lambda$.
\subsection{Reactive Jammer}
\label{subsec:jammer}
In this work, we consider a powerful reactive jammer\footnote{Our system can be extended to the case with multiple jammers that can cooperate to attack the channel.} that can attack the channel when it detects active transmissions from the transmitter. Intuitively, the reactive jammer can detect if the transmitter is transmitting data or not by using common signal detection techniques, e.g., energy detection, which is the most common channel sensing method with very small sensing time~\cite{Fang2016Wireless},~\cite{Xu2005The}. As soon as the transmitter transmits data, the jammer can attack the channel by transmitting strong signals to the channel to reduce the effective signal-to-interference-plus-noise ratio (SINR) at the receiver. In particular, under the jammer attack, the SINR at the receiver can be formally expressed by~\cite{Hanawal2016Joint},~\cite{Firouzbakht2012RA}:
\begin{equation}
\label{eq:SIRN}
\theta = \frac{P^\mathrm{R}}{\phi P^\mathrm{J} + \rho^2},
\end{equation}
where $P^\mathrm{R}$ is the power transmitted from the transmitter received at the receiver, $\phi P^\mathrm{J}$ denotes the jamming power received at the receiver with $0 \le \phi \le 1$ expresses the attenuation factor, and $\rho^2$ is the variance of additive white Gaussian noise. We denote $P_{\mathrm{avg}}$ as the time-average power constraint of the jammer and $P_{\max}$ as the peak jamming power, i.e., $P_{\mathrm{avg}} \leq P_{\max}$~\cite{Firouzbakht2012RA}. 

In this work, once detecting activities of the transmitter on the target channel, the reactive jammer can attack the legitimate system at various power levels with different attack probabilities. This strategy is much more intelligent and beneficial for the jammer than the fixed strategy, i.e., always attack at the same power level, as the jammer can adjust its attack power levels based on the activities of the transmitter, e.g., actively transmit, use the RA technique or stay idle. Let $\mathbf{P}_{\mathrm{J}} = \{P^{\mathrm{J}}_0, \ldots, P^{\mathrm{J}}_n, \ldots, P^{\mathrm{J}}_N\}$ denote a set of discrete power levels the jammer can use to attack the channel. We denote $\mathbf{x} \triangleq (x_0,\ldots, x_n, \ldots, x_N)$ as an attack probability vector. In each time slot, when the transmitter actively transmits data on the target channel, the jammer can attack the channel with power level $P^{\mathrm{J}}_n$ if the average power constraint is satisfied. Denote $\mathbf{J}_s$ as the attack strategy of the jammer. We have
\begin{equation}
\begin{aligned}
\label{eq:attack_strategy}
\mathbf{J}_s \triangleq \Big\{(x_0,\ldots, x_n, \ldots, x_N): \sum_{n=0}^{N} x_n = 1, x_n \in [0,1], \forall n \in \{0,\ldots,N\}, \mathbf{x} \mathbf{P}_{\mathrm{J}}^\top \leq P_{\mathrm{avg}}\Big\}.
\end{aligned}
\end{equation}
In this work, we consider a scenario in which the location of jammer (as well as its jamming attack capability) is unknown. Hence, the conventional methods to analyze and estimate the received signals at the receiver may not be feasible to implement in this case because they may require accurate channel estimations as well as the location of jammer for calculating. As a result, in this work, we aim to develop an intelligent framework to deal with this problem. Specifically, this framework is empowered by advanced reinforcement learning algorithms in order to explore and learn from surrounding environment and the jammer, and thereby making appropriate actions to maximize its performance. In this way, the information about the environment and the jammer, e.g., the number of packets can be transmitted, the number of energy units can be harvested, and the jammer's strategy and location, can be gradually captured and estimated through the learning process.
\subsection{Deception Strategy}
In this paper, we propose an intelligent deception mechanism that allows the system to not only undermine the jammer's attack efficiency but also leverage the jamming signals to improve the system performance. In particular, at the beginning of each time slot, the transmitter can lure the jammer by actively transmitting ``fake'' signals for a short period of time\footnote{It is noted that the deception time must be higher than the detection time of the jammer. The transmitter can observe the activities of jammer during the learning process, and then determine the optimal deception time. Determining the optimal deception time is out of scope of this paper.}. After that, the transmitter listens the channel to detect activities of the jammer by using common signal detection techniques~\cite{Fang2016Wireless},~\cite{Xu2005The}. If the jammer attacks the channel, the transmitter can leverage the jamming signals to support its operations. This strategy is very beneficial for low-power systems even in dealing with a very powerful reactive jammer. The reason is that if the jammer often attacks the channel at high power levels, the jammer will unintentionally provide an abundant energy resource to supply for the legitimate system. Note that the ``fake'' signals can be generated and transmitted like the actual signals (but transmitted in a short period of time instead of the whole time frame as the actual signals). In this case, after the deception period, if the jammer attacks the channel, the transmitter can leverage the jamming signals to harvest energy and backscatter data. Note that to save the energy for the transmitter, the transmitter may not need to perform complex modulation/encoding techniques when transmitting ``fake'' signals.

In the system under considerations, we assume that time is slotted, and at the beginning of a time slot, the transmitter can choose to actively transmit data or perform deception strategy. If the transmitter chooses to actively transmit data and the jammer does not attack the channel (as illustrated in Fig.~\ref{Fig.system_model}(a)), the transmitter can successfully transmit $\widehat{d}_\mathrm{a}$ packets to the receiver. If the jammer attacks the channel, the transmitter cannot actively transmit packets to the receiver as shown in Fig.~\ref{Fig.system_model}(b). We assume that the transmitter requires $e_\mathrm{r}$ energy units to transmit one packet. On the other hand, if the transmitter performs the deception, it will first transmit signals on the channel for a short period. After that, the transmitter listens the channel to detect activities of the jammer. We denote $e_\mathrm{f}$ as the total amount of energy that the transmitter needs to perform deception (including the sensing process). If the jammer stays idle after the deception, it can actively transmit $\widehat{d}_\mathrm{de}$ packets to the receiver in the rest of the time slot ($\widehat{d}_\mathrm{de} < \widehat{d}_\mathrm{a}$) as illustrated in Fig.~\ref{Fig.system_model}(c). In contrast, if the jammer attacks the channel, the transmitter can choose one of three actions: (i) use the RA technology to reduce the transmission rate, (ii) harvest energy from the jamming signals, or (iii) backscatter the jamming signals as illustrated in Fig.~\ref{Fig.system_model}(d), Fig.~\ref{Fig.system_model}(e), and Fig.~\ref{Fig.system_model}(f), respectively. Note that although the jamming signals are noise-like signals~\cite{Lichtman2012Reinforcement}, the transmitter still can leverage the jamming signals for its operations. In particular, as demonstrated in~\cite{Gu02017Exploiting} and~\cite{Rezgui2019Mitigating}, the transmitter can harvest energy from the jamming signals as the existing energy harvesting circuit is able to harvest from the noise-like signals. Moreover, with the ambient backscatter technology, the transmitter can backscatter the noise-like signals, as studied in~\cite{Guo2019Nocoherent} and~\cite{Zhang2019Constellation}.
\subsubsection{Rate Adaptation}
By using the rate adaptation technique, the transmitter can transmit data at rate $r_m$ when the jammer attacks the channel with jamming power $P^{\mathrm{J}}_\mathrm{n}$. We then denote $\mathbf{r}=\{r_1, \ldots, r_m, \ldots,  r_M\}$ as the vector of $M$ transmission rates that the transmitter can use under jamming attacks. With each rate $r_m$, the transmitter can transmit $\widehat{d}^{\mathrm{r}}_m$ packets to the receiver. We define $\gamma_m$ as the lowest SINR value at which the receiver can successfully decode packets sent at rate $r_m$. The higher transmission rate requires the higher value of SINR at the receiver~\cite{Hanawal2016Joint}. Thus, for $m=1, \ldots, M$, when $\gamma_{m-1} \leq \theta < \gamma_m $, the receiver cannot decode packets that are transmitted at rates higher than $r_{m-1}$~\cite{Hanawal2016Joint}. As mentioned, by using common signal detection techniques, the transmitter can observe the jammer's activities after performing the deception. As such, the transmitter can choose an appropriate rate to transmit data based on the estimated jamming power. However, if the transmitter fails to detect the attack power level of the jammer (due to the miss detection), the transmitted packets will be lost. We then denote $p_\mathrm{miss}$ as the miss detection probability of the transmitter in detecting attack power levels of the jammer.
\subsubsection{RF Energy Harvesting}
After performing the deception mechanism, the transmitter can harvest RF energy from the jamming signals if the jammer attacks the channel. We denote $e^{\mathrm{J}}_n$ as the number of harvested energy when the jammer attacks with power level $P^{\mathrm{J}}_\mathrm{n}$. The harvested energy is stored in the energy storage to support for future deception and actual transmission activities~\cite{Huynh2018Survey}. We then denote $\mathbf{e}=\{e^{\mathrm{J}}_0,\ldots, e^{\mathrm{J}}_n, \ldots, e^{\mathrm{J}}_N\}$ as the set of harvested energy that the transmitter can harvest according to the jamming power levels. Intuitively, the amount of harvested energy increases with the attack power level of the jammer. This proportional relationship can be observed clearly through the Friis equation~\cite{Balanis2012Antenna} in which the amount of harvested energy can be expressed by a linear function of transmission power of the jammer.
\subsubsection{Ambient Backscatter Communications}
\begin{figure*}[!]
	\centering
	\includegraphics[scale=0.2]{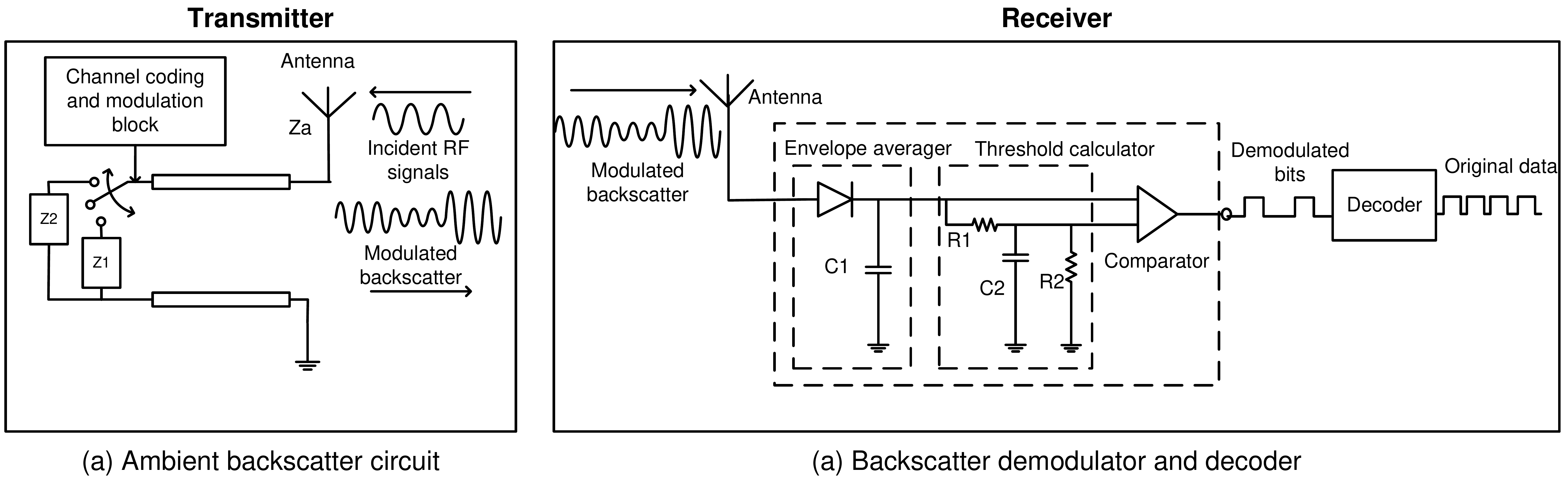}
	\caption{Ambient backscatter circuit~\cite{Liu2013Ambient,Huynh2018Survey}.}
	\label{Fig.ambientBackscatter}
\end{figure*}

The transmitter can also backscatter the jamming signals to transmit data to the receiver by using the ambient backscatter circuit as shown in Fig.~\ref{Fig.ambientBackscatter}(a)~\cite{Liu2013Ambient}.
In particular, by switching between two loads $Z_1$ and $Z_2$ with an RF switch, e.g., ADG902, the transmitter can switch between two states: (i) reflecting and (ii) non-reflecting. At the non-reflecting state, all the RF signals (i.e., the jamming signals in this paper) will be absorbed, and thus this state represents bits `0'. Otherwise, at the reflecting state, all the RF signals will be reflected to the receiver, and thus this state represents bits `1'. In this way, the transmitter can backscatter data to the receiver without generating active signals. It is worth noting that at the non-reflecting state, the transmitter still can harvest RF energy~\cite{Liu2013Ambient}. However, the amount of harvested energy in this case is not significant and only enough to support the operations of the ambient backscatter mode. It is worth noting that the more power the RF source uses to transmit signals (e.g., the reactive jammer generates high-power interference to attack the channel), the better performance of backscatter communications is~\cite{Liu2013Ambient,Boyer2014Backscatter,Kimionis2012Bistatic}. The reason is that the backscattered signal received at the receiver is stronger, resulting in high detection/decoding performance. Several ambient backscatter prototypes have been introduced in the literature with backscatter rates up to few Mbps~\cite{Huynh2018Survey}. Thus, using ambient backscatter to leverage the strong jamming signals is a very promising solution.

By using the ambient backscatter technology, the transmitter can transmit $\widehat{d}_n^\mathrm{J}$ packets to the receiver when the jammer attacks the channel with power level $P^\mathrm{J}_n$ after deception. We denote $\widehat{\mathbf{d}}=\{\widehat{d}^{\mathrm{J}}_0,\ldots,\widehat{d}^{\mathrm{J}}_n, \ldots,\widehat{d}^{\mathrm{J}}_N\}$ as the set of packets that the transmitter can backscatter to the receiver corresponding to the jamming power levels of the jammer. Note that the backscatter rate depends on the hardware configuration, i.e., the values of the RC circuit elements in Fig.~\ref{Fig.ambientBackscatter}(b)~\cite{Liu2013Ambient}. Thus, in this work, we consider that the backscatter rate is fixed at a maximum rate of $d_{\max}$ packets. If the maximum number of packets that can be backscattered at jamming power $P^\mathrm{J}_n$ is lower than $d_{\max}$, ($d_{max} - \widehat{d}^{\mathrm{J}}_n$) packets will be lost as the jamming signals is not strong enough to support the transmitter to transmit all $d_{\max}$ packets.

To decode the backscattered signals, there are two approaches: (i) using an analog-to-digital converter (ADC) and (ii) using the averaging mechanism~\cite{Liu2013Ambient},~\cite{Huynh2018Survey}. However, as the ADC consumes a significant amount of energy, the averaging mechanism is usually adopted in the literature (especially in IoT networks) to allow the receiver to decode the backscattered signals by using analog components as illustrated in Fig.~\ref{Fig.ambientBackscatter}(b). The key idea of the averaging mechanism is using the envelope-averaging circuit to smooth the backscattered signals received at the receiver. Then, the compute-threshold circuit is used to determine the voltage between the high level and low level of the smoothed signals. Finally, the output bit is decided by comparing this voltage with a threshold (predefined according to scenarios and circuits). Interested readers can find more information about the hardware design and decoding algorithms in~\cite{Liu2013Ambient}.

Note that there are some scenarios where the ambient backscatter technology may not work well under jamming attacks. However, in these scenarios, the transmitter can choose other communication methods, e.g., harvest-then transmit methods, to maximize its performance. For example, when the jammer is placed far from the receiver, the backscatter communication method can be used if its efficiency is greater than that that of the harvest-then-transmit or rate adaption method. However, in the cases if the jammer is placed too close to the receiver, the receiver may not be able to decode the backscattered signals or direct signals transmitted by using the RA technology as the jamming signals received at the receiver is much stronger than these signals. In this case, the transmitter can choose to harvest energy from the jamming signals and wait until the jammer does not attack the channel to actively transmit data to the receiver. Note that backscatter communication is just one option for the transmitter to choose from, and through the learning process, the transmitter can gradually learn from the jammer activities and adapt its communication strategy accordingly.

\subsection{Jammer Performance Analysis}
\label{sec:jammeranalys}
To theoretically elaborate the efficiency of proposed deception mechanism, in the following, we evaluate the jammer utility under the transmitter's deception strategy. At the beginning of each time slot, if the transmitter actively transmits actual data, and if the jammer attacks the channel, all packets transmitted by the transmitter are lost. In this case, the jammer can receive a reward of $0 \le d_\mathrm{a} \leq \widehat{d}_\mathrm{a}$ (corresponding to the number of dropped packets). Thus, the jammer's utility function for this case can be  expressed as follows:
\begin{equation}
	U_1^\mathrm{J} 	=\!\!	\left\{	\begin{array}{ll}
		\!\!d_\mathrm{a}\!\!\!	&\mbox{if the transmitter transmits actual data and the jammer attacks the channel}. \\
		\!\!-d_\mathrm{a} \!\!\! &\mbox{if the transmitter transmits actual data and the jammer does not attack the channel.}
	\end{array}	\right.
\end{equation}
If the transmitter chooses to use $e_\mathrm{f}$ units of energy to perform the deception, the jammer will get a reward of $\frac{e_\mathrm{f}}{e_\mathrm{r}}$. Here, $\frac{e_\mathrm{f}}{e_\mathrm{r}}$ can be interpreted as the potential number of packets that the transmitter can transmit without performing deception. If the jammer decides to attack the channels, the transmitter can leverage the jamming signals to harvest $e^\mathrm{J}_\mathrm{n}$ units of energy or backscatter $0 \le d^\mathrm{J}_\mathrm{n} \leq \widehat{d}^\mathrm{J}_\mathrm{n}$ packets or using the rate adaptation to transmit $0 \le d^{\mathrm{r}}_m \leq \widehat{d}^{\mathrm{r}}_m$ packets. Note that the harvested energy can be used to actively transmit data or perform deception actions later. Thus, the penalty for the jammer if the transmitter harvests energy from the jamming signals can be expressed by $-\frac{e^\mathrm{J}_\mathrm{n}}{e_\mathrm{r}}$, i.e., the number of potential packets which the transmitter can transmit from the harvested energy. Similarly, we can denote $-d^\mathrm{J}_\mathrm{n}$ and $-d^{\mathrm{r}}_m$ to be the penalties for the jammer if the transmitter uses backscatter and rate adaptation techniques to transmit data, respectively. Hence, the utility function of the jammer in this case can be expressed as follows:
\begin{equation}
	U_2^\mathrm{J} 	=	\left\{	\begin{array}{ll}
		\frac{e_\mathrm{f}}{e_\mathrm{r}} -\frac{e^\mathrm{J}_\mathrm{n}}{e_\mathrm{r}} &	\mbox{if the transmitter harvests energy from the jamming signals}, \\
		\frac{e_\mathrm{f}}{e_\mathrm{r}} -d^\mathrm{J}_\mathrm{n} &	\mbox{if the transmitter backscatters data through the jamming signals}, \\
		\frac{e_\mathrm{f}}{e_\mathrm{r}} -d^{\mathrm{r}}_m &	\mbox{if the transmitter adapts its transmission rate},\\
		\frac{e_\mathrm{f}}{e_\mathrm{r}}  &	\mbox{if the transmitter stays idle}.
	\end{array}	\right.
\end{equation}
Finally, if the transmitter performs deception, but the jammer does not attack the channel, then the transmitter can actively transmit data in the rest of the time slot. In this case, the transmitter will waste $e_\mathrm{f}$ units of energy for the deception action, but it can successfully transmit $0 \le d_\mathrm{de} \leq \widehat{d}_\mathrm{de}$ packets to the receiver. Thus, we can derive the utility function of the jammer in this case as follows:
\begin{equation}
U_3^\mathrm{J} 	=	\left\{	\begin{array}{ll}
\frac{e_\mathrm{f}}{e_\mathrm{r}} - d_\mathrm{de}&	\mbox{if the transmitter transmits data}, \\
\frac{e_\mathrm{f}}{e_\mathrm{r}} & \mbox{if the transmitter stays idle.}
\end{array}	\right.
\end{equation}
Then, we derive the jammer's expected overall utility as follows:
\begin{equation}
\label{Eq:utilityJammer}
U = U_1^\mathrm{J} + U_2^\mathrm{J} + U_3^\mathrm{J}.
\end{equation}
In~(\ref{Eq:utilityJammer}), it can be observed that if the jammer attacks the channel at high frequency and at the same time the transmitter often performs deception strategy, then the efficiency of jamming attack will be significantly reduced. However, if the jammer does not often attack the channel and the deception probability is high, then the deception strategy is not effective. As a result, in order to maximize the performance for the system, the transmitter needs to know the jammer's strategy, e.g., power levels and frequency of attacks, in advance. Unfortunately, this information is usually unknown by the transmitter in advance. Thus, in this paper, we propose reinforcement learning approaches to enable the transmitter to deal with the uncertainty and dynamic of the jammer and the environment by learning from real-time interactions.

Note that the amount of harvested energy and the number of backscattered/transmitted packets, i.e., $d_\mathrm{a}$, $d_\mathrm{de}$, $d^\mathrm{J}_\mathrm{n}$, $e^\mathrm{J}_\mathrm{n}$, and $d^{\mathrm{r}}_m$, can be observed after interacting with the jammer. Thus, our proposed solutions do not require this explicit information in advance. Instead, the learning algorithms learn these values and find the optimal policy for the transmitter. For example, in the case if the jammer often attacks the channel at low power levels, the amount of harvested energy and the number of backscattered packets could be low. As such, our proposed reinforcement learning algorithms can learn and find the optimal policy to guide the transmitter to choose the best actions, e.g., rate adaptation instead of backscattering, to maximize the system performance.
\section{Problem Formulation}
\label{Sec:prob}
\begin{figure}[!]
	\centering
	\includegraphics[scale=0.27]{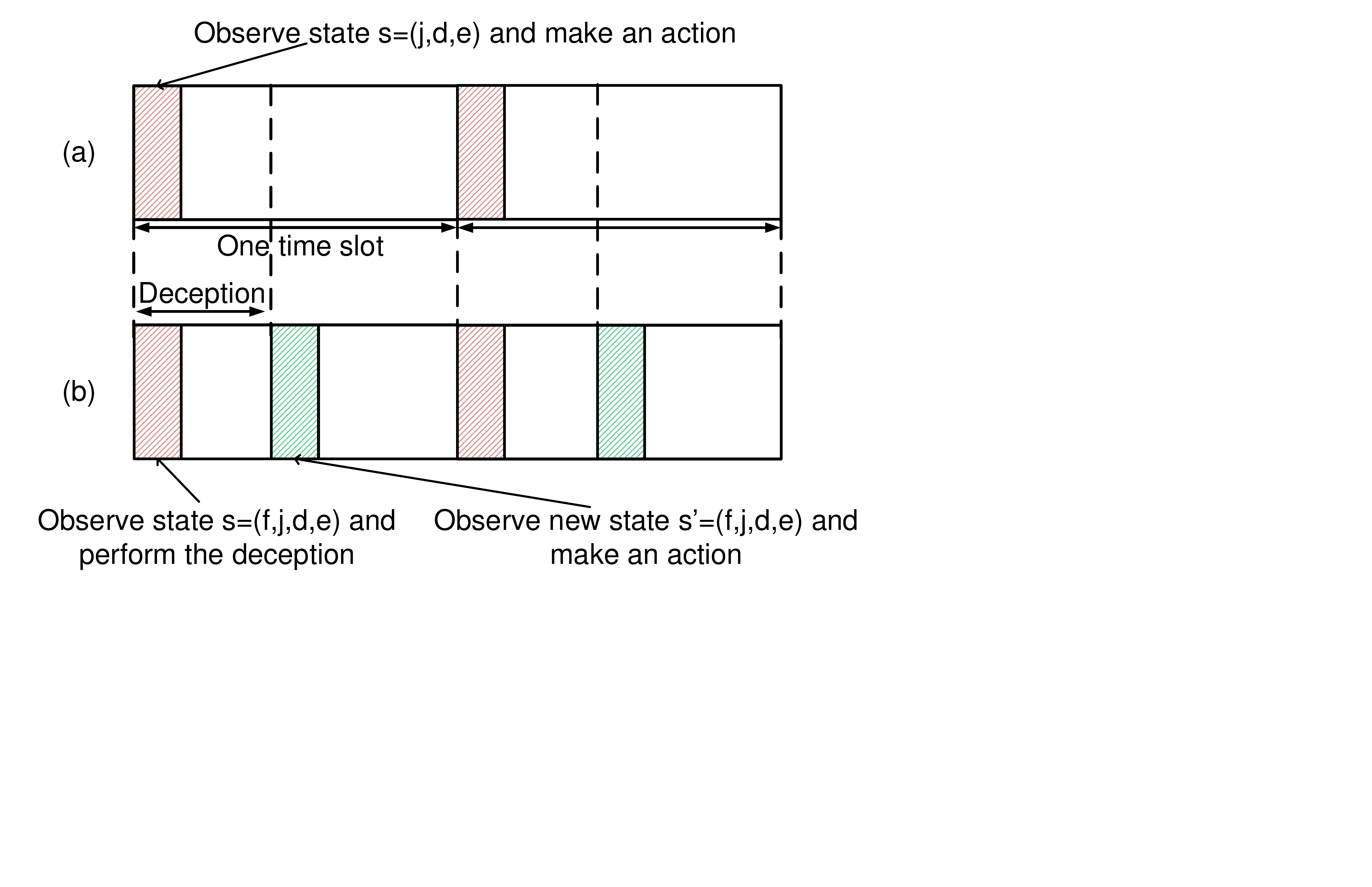}
	\caption{Decision epoch of (a) conventional MDP and (b) proposed MDP.}
	\label{Fig.decision_epoch}
\end{figure}

To learn from and adapt with the jammer's behaviors as well as the uncertainty of the environment, we adopt the Markov decision process (MDP) framework to formulate the optimization problem. The MDP is defined by a tuple $<\mathcal{S}, \mathcal{A}, r>$ where $\mathcal{S}$ is the state space, $\mathcal{A}$ is the action space, and $r$ is the immediate reward function of the system. For a conventional MDP process, at the beginning of a time slot, the transmitter observes the current system state, e.g., data, energy and channel states, performs an action, e.g., active data transmission or deception, and observes the results in the end of the time slot, e.g., packets are successfully transmitted or dropped. However, this conventional process is not appropriate to adopt in our model to defeat the reactive jammer. The reason is that the reactive jammer only attacks channel if it detects activities of the transmitter on the channel, and thus at the beginning of a time slot, the channel state is always idle. Second, for the conventional MDP process with only one decision epoch, after the transmitter performs deception and the jammer attacks the channel, we only can undermine the jammer's power, but cannot leverage jamming signals for enhancing the system performance. We thus propose a new MDP model with two decision epochs, i.e., one at the beginning of a time slot and another is after the deception period as illustrated in Fig.~\ref{Fig.decision_epoch}. To be more specific, at the first decision epoch, i.e., at the beginning of a time slot, the transmitter observes the current system state, including data queue, energy queue, deception and channel states, and makes an action, e.g., deception or actively transmit actual data. Then, at the beginning of the second decision epoch, i.e., right after the deception period, the transmitter observes the new states, e.g., whether the jammer attacks the channel or not, and then makes an action, e.g., backscatter data or harvest energy by leveraging the jamming signals when the jammer attacks the channel. In this way, we can not only undermine the jammer's power, but also utilize its power for improving the system performance.
\subsection{State Space}
The state space of the system can be defined as follows:
\begin{equation}
\begin{aligned}
\mathcal{S} \triangleq \Big\{ (f, j, d,e) \mid f \in \{0,1\}; j \in \{0, 1\}; d \in \{0,\ldots, D\}; e \in \{0,\ldots, E \} \Big\} \setminus \{0,1,d,e\},
\end{aligned}
\end{equation}
where $d$ and $e$ represent the number of packets in the data queue and the number of energy units in the energy storage of the transmitter, respectively. $f$ represents the deception strategy of transmitter, i.e., $f=1$ when the deception is performed and $f=0$ otherwise. Note that $f$ is always 0 at the first epoch, but could be 0 or 1 in the second epoch of a time slot. $j$ represents the state of the jammer, i.e., $j = 1$ when the jammer attacks the channel and $j = 0$ otherwise. Note that at the first epoch of a time slot, $j$ is always 0. However, after the deception is made, $j$ could be 0 or 1. Moreover, at the second epoch, the jammer only attacks the channel if the transmitter performs the deception at the first epoch. Thus, the system state space does not include state $s = \{0,1,d,e\}$. Finally, we can define the state of the system as $s = (f,j,d,e) \in \mathcal{S}$.
\subsection{Action Space}
We denote $\mathcal{A} \triangleq \{a:a \in \{0, 1,\ldots, 4+m\} \}$ as the action space of the transmitter in which $a=1$ represents the action of performing the deception. At the beginning of a time slot, if the transmitter has enough energy (i.e., $e \geq e_\mathrm{f}$), it can choose to perform the deception. In this case, we have the following actions:
\begin{equation}
	a 	=	\left\{	\begin{array}{ll}
		2,	&	\mbox{the transmitter transmits data if $f=1$, $j=0$, $d > 0$, and $e > e_\mathrm{r}$}, \\
		3,	&	\mbox{the transmitter harvests energy if  $f=1$,$j=1$, and $e<E$},\\
		4,	&	\mbox{the transmitter backscatters data if $f=1$,$j=1$, and $d > 0$},\\
		4+m,&  	\mbox{the transmitter adapts its transmission rate to $r_m$ if $f=1$, $j=1$, $d>0$, and $e > e_\mathrm{r}$,}\\
		0, & \mbox{the transmitter stays idle.}
	\end{array}	\right.
\end{equation}
In particular, after performing the deception, the transmitter listens to the channel. If the jammer stays idle, i.e., $j=0$, the data queue is not empty, and there is enough energy, the transmitter can choose to actively transmit data to the receiver in the rest of the time slot. If the jammer attacks the channels, i.e., $j=1$, the transmitter can choose to harvest energy from the jamming signals in the rest of the time slot. If the data queue is not empty, the transmitter can choose to backscatter data to the receiver in the rest of the time slot. Additionally, the transmitter can choose to reduce its data rate to transmit data to the receiver if it has data in the data queue and has sufficient energy for active transmissions in the energy storage. Otherwise, the transmitter can choose to stay idle for the rest of the time slot.

\begin{figure}[!]
	\centering
	\includegraphics[scale=0.37]{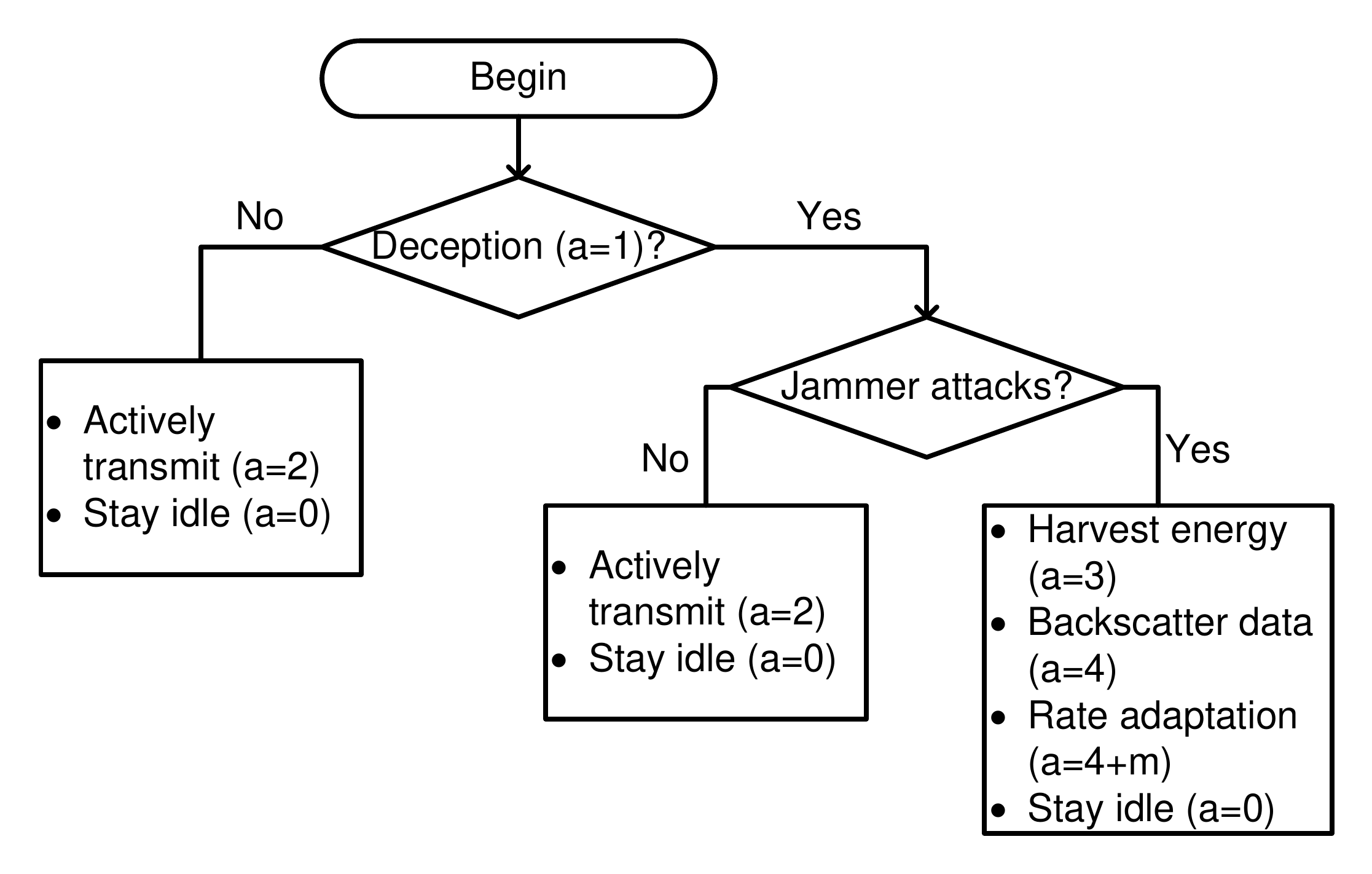}
	\caption{Flowchart to express actions of transmitter.}
	\label{Fig.flowchartDeception}
\end{figure}

If the transmitter decides not to perform the deception at the beginning of the time slot, we have the following actions:
\begin{equation}
	a 	=	\left\{	\begin{array}{ll}
		2,	&	\mbox{the transmitter actively transmits data}\\
		&\mbox{if $d > 0$ and $e > e_\mathrm{r}$},	\\
		0,	&	\mbox{the transmitter stays idle.}	\\
	\end{array}	\right.
\end{equation}
Specifically, in this case, the transmitter can choose to actively transmit data if it has data in the data queue and has enough energy in the energy storage for active transmissions. Alternatively, the transmitter can decide to stay idle in this time slot. The flowchart of the transmitter's actions is illustrated in Fig.~\ref{Fig.flowchartDeception}. Note that the rules in Fig.~\ref{Fig.flowchartDeception} are used to avoid infeasible actions that lead to unreachable states in the Markov decision processes. Thus, using these rules can not only ensure the properness of the learning process, but also make the agent learning more effectively because the number of actions at each state can be reduced. To the end, the rules presented in Fig.~\ref{Fig.flowchartDeception} do not limit the agent to learn new policies. Instead, they are used to control the agent to learn properly.

\subsection{Immediate Reward}
In this work, the reward function is denoted by the number of packets that the transmitter can successfully transmit to the receiver. Hence, the immediate reward over one time slot after the transmitter takes an action $a$ at state $s$ can be defined as follows:
\begin{equation}
\label{eq:reward}
r(s,a)	=	\left\{	\begin{array}{ll}
d_\mathrm{a},	& \mbox{if $a=2$ and $f = 0$},\\
d_\mathrm{de} & \mbox{if $a=2$ and $f=1$},	\\
d_\mathrm{n}^\mathrm{J},	& \mbox{if $a=4$},	\\
d^\mathrm{r}_m, & \mbox{if $a=4+m$},\\
0	,						&	\mbox{otherwise}	. \end{array}	\right.
\end{equation}
In~(\ref{eq:reward}), if the transmitter chooses to actively transmit data at the beginning of a time slot and the jammer does not attack the channel, it can successfully transmit $d_\mathrm{a} \leq \widehat{d}_\mathrm{a}$ packets to the receiver. If the transmitter performs deception and the jammer stays idle, the transmitter can transmit $d_\mathrm{de} \leq \widehat{d}_\mathrm{de}$ packets to the receiver. However, if the transmitter performs deception and the jammer attacks the channel, the transmitter can backscatter $d_\mathrm{n}^\mathrm{J} \leq \widehat{d}_\mathrm{n}^\mathrm{J}$ packets or transmit $d^\mathrm{r}_m \leq \widehat{d}^\mathrm{r}_m$ packets (using rate adaption technique) to the receiver. Note that the harvested energy can be used to actively transmit data to the receiver and/or to perform the deception mechanism. Thus, the ``actual'' reward can be finally measured based on the number of packets successfully transmitted. This also aligns with the main goal of the legitimate system, i.e., maximize its throughput. Finally, the immediate reward is $0$ if the transmitter cannot successfully transmit any packet to the receiver. Is is worth noting that after performing an action, the transmitter observes its reward, i.e., the number of packets successfully transmitted based on ACK messages sent from the receiver. In other words, $d_\mathrm{a}$, $d_\mathrm{de}$, $d_\mathrm{n}^\mathrm{J}$, and $d^\mathrm{r}_m$ are the actual number of packets received at the receiver. As such, the reward function captures the overall path between the receiver and the transmitter, e.g., BER, fading, or end-to-end SNR. Note that the feedback transmitted from the receiver to the transmitter can also be disrupted by the jamming signals. Nevertheless, we can mitigate this problem by using the same anti-jamming strategy as that of the transmitter. In particular, like the transmitter, the receiver can use the ambient backscatter communication technology to backscatter the feedback to the transmitter when the jammer attacks the channel.
\subsection{Optimization Formulation}
In this work, we aim to obtain the optimal defense policy to maximize the average long-term throughput of the system, denoted by $\pi^*:\mathcal{S} \rightarrow \mathcal{A}$. In particular, the optimal policy is a mapping from a given state to an optimal action. The optimization problem can be formulated as follows:
\begin{eqnarray}
\label{eq:average_reward}
\max_\pi	& &	{\mathcal{R}}(\pi)	=	\lim_{T \rightarrow \infty} \frac{1}{T} \sum_{k=1}^{T} {\mathbb{E}} \left( r_k (s_k, \pi(s_k)) \right),	\label{eq:cmdp_obj}
\end{eqnarray}
where $r_k (s_k, \pi(s_k))$ denotes the immediate reward at time step $k$ given policy $\pi$ and ${\mathcal{R}}(\pi)$ is the average reward with policy $\pi$. In Theorem~\ref{theo:limitexists}, we show that the average throughput $\mathcal{R}(\pi)$ is well defined and does not depend on the initial state.
\begin{theorem}
	\label{theo:limitexists}
	For every $\pi$, the average throughput $\mathcal{R}(\pi)$ is well defined and does not depend on the initial state.
\end{theorem}
The proof of Theorem~\ref{theo:limitexists} is provided in Appendix~\ref{appendix:limitexist}.

\section{Optimal Defense Strategy with Reinforcement Learning Algorithms}
\label{sec:QDeepQ}
\subsection{Q-Learning based Deception Strategy}
This section presents the Q-learning algorithm to obtain the optimal defense policy for the transmitter. In particular, this algorithm does not require the information about the jammer in advance. Instead, the Q-learning algorithm can explore the environment and learn from its experiences. Theoretically, the Q-learning algorithm is ensured to converge to the optimal policy with probability one as proved in~\cite{Watkins1992QLearning}. In the following, we present the key idea behind the Q-learning algorithm.

We denote $\pi^*:\mathcal{S} \rightarrow \mathcal{A}$ as the optimal defense policy, which is a mapping from system states to their corresponding actions, for the transmitter under the jamming attacks. For each policy $\pi$, the expected value function $\mathcal{V}^\pi(s): \mathcal{S} \rightarrow \mathbb{R}$ can be expressed as follows:
\begin{equation}
\begin{aligned}
\mathcal{V}^\pi(s) = \mathbb{E}_\pi \Big [ \sum_{t=0}^{\infty} \gamma^t r_t(s_t, a_t)|s_0=s\Big ] =\mathbb{E}_\pi\Big [ r_t(s_t, a_t) + \gamma\mathcal{V}^\pi(s_{t+1})|s_0=s\Big ],
\end{aligned}
\end{equation}
where $r_t(s_t, a_t)$ denotes the immediate reward after taking action $a_t$ at state $s_t$. $\gamma$ is the discount factor that denotes the importance of the long-term reward~\cite{Watkins1992QLearning}. To find the optimal policy $\pi^*$, the optimal action at each state can be found by using the following optimal value function.
\begin{equation}
\label{eq:V}
\mathcal{V}^*(s) = \max_{a} \Big \{ \mathbb{E}_\pi[r_t(s_t, a_t)+ \gamma\mathcal{V}^\pi(s_{t+1})] \Big\} ,\quad \forall s \in \mathcal{S}.
\end{equation}
Thus, the optimal Q-functions for all pairs of states and actions are derived as follows:
\begin{equation}
\label{eq:Q}
\mathcal{Q}^*(s,a) \triangleq r_t(s_t, a_t) + \gamma\mathbb{E}_\pi[\mathcal{V}^\pi(s_{t+1})] , \quad \forall s \in \mathcal{S}.
\end{equation}

From (\ref{eq:V}) and (\ref{eq:Q}), the optimal value function $\mathcal{V}^*(s)$ can be expresses as $\mathcal{V}^*(s) = \max_{a} \{ \mathcal{Q}^*(s,a)\}$. It can be observed that the problem can be transformed to finding the optimal Q-value $\mathcal{Q}^*(s,a)$ for each state-action pair $(s,a)$ by iteratively taking samples as follows:
\begin{equation}
\label{Eq:qfunction}
\begin{aligned}
\mathcal{Q}_{t+1}(s_t,a_t) = \mathcal{Q}_t(s_t,a_t) + \tau_t \Big [ r_t(s_t, a_t) + \gamma\max_{a_{t+1}} \mathcal{Q}_t(s_{t+1}, a_{t+1})- \mathcal{Q}_t(s_t,a_t)\Big ].
\end{aligned}
\end{equation}
The principle of (\ref{Eq:qfunction}) is finding the temporal difference between the current estimated Q-value $\mathcal{Q}_t(s_t,a_t)$ and the target Q-value $r_t(s_t, a_t) + \gamma \max_{a_{t+1}} \mathcal{Q}_t(s_{t+1}, a_{t+1})$. By updating the Q-table based on (\ref{Eq:qfunction}), the algorithm can gradually converge to the optimal policy. The learning rate $\tau_t$ presents the influence of the new experience on the current estimated Q-value. In practice, the learning rate can be adjusted during the training process. However, to ensure that the Q-learning algorithm always converges to the optimal policy, the learning rate must be nonnegative, deterministic, and follows the following rules~\cite{Watkins1992QLearning}:
\begin{equation}
\label{Eq:rules}
\tau_t \in [0,1), \sum_{t=1}^{\infty}\tau_t = \infty, \mbox{ and } \sum_{t=1}^{\infty} ( \tau_t  )^{2} < \infty.
\end{equation}

Based on (\ref{Eq:qfunction}), the algorithm updates the Q-values for all state-action pairs. In particular, based on the $\epsilon$-greedy algorithm, at the current state $s_t$, the algorithm selects a random action with probability $\epsilon$ and selects an action that maximizes the Q-value function with probability $1-\epsilon$. Next, the algorithm performs the selected action and observes the immediate reward and the next state of the environment. These observations are then used to update the Q-table based on~(\ref{Eq:qfunction}). After a finite number of iterations, the algorithm will obtain the optimal defense policy for the system~\cite{Watkins1992QLearning}. Nevertheless, the Q-learning based algorithms are well-known for their slow-convergence, especially in complicated systems with high-dimensional state and actions spaces. To deal with this problem, in the following, we propose the deep dueling algorithm to allow the transmitter to obtain the optimal policy with a much faster convergence rate by leveraging the deep Q-learning and novel dueling architecture.

\subsection{Deep Q-Learning based Deception Strategy}
This section presents the deep Q-learning algorithm~\cite{Mnih2015Human} to improve the learning rate of the Q-learning algorithm. Different from the Q-learning algorithm, the deep Q-learning algorithm estimate the values of $\mathcal{Q}^*(s,a)$ by using a deep neural network architecture instead of the Q-table. The deep neural network can handle the high-dimensional system states very well, and thus improve the convergence rate of the learning process. However, as stated in~\cite{Mnih2015Human}, using a nonlinear function approximator may lead to the instability and divergence of the deep Q-learning algorithm. This is stemmed from the fact that when the Q-values are slightly updated, the policy may be significantly changed, and therefore the correlations between the estimated Q-values and the target Q-values (i.e., $r+\gamma \max_{a} \mathcal{Q}(s,a)$) as well as the data distribution are greatly affected. Therefore, in this paper, we adopt two mechanisms to address this drawback as follows:

\begin{itemize}
	\item \textit{Experience replay mechanism:} All transitions $(s_t, a_t, r_t, s_{t+1})$, i.e., experiences, of the algorithm are stored in a memory pool $\mathbf{D}$. The algorithm then randomly chooses a number of samples from the replay memory and feds to the deep neural network for training. As such, the previous experiences can be efficiently learned many times to improve the stability of the learning process.
	
	\item \textit{Quasi-static target Q-network:} During the training process, the Q-values for each pair of state and action will be changed. Thus, the algorithm may not be stable if a constantly shifting set of values is used to update the Q-network. To solve this problem, we use the quasi-static target network method to improve the stability of the algorithm. In particular, we implement a target Q-network and frequently but slowly update its network parameters with the Q-network parameters. The target Q-values are then obtained based on the target Q-network. In this way, the correlations between the estimated Q-values and the target Q-values can be eliminated, resulting in good learning performance.
\end{itemize}
Moreover, in this work, we smartly design the feature set of the deep neural network to further improve the learning performance. In particular, the feature set consists of the state of the jammer, the state of the deception mechanism, and the states of the data queue and the energy queue. The deep neural network then takes these features as it inputs for training process. As a result, all properties of the system states are well trained by the neural network, and thus improving the learning performance.
\begin{algorithm}
	\caption{Deep Q-learning Based Anti-jamming Algorithm}
	\label{deepqlearning}
	\begin{algorithmic}[1]
		\State Initialize replay memory $\mathbf{D}$ to capacity $\mathcal{D}$.
		\State Initialize the Q-network $\mathcal{Q}$ with random weights $\theta$.
		\State Initialize the target Q-network $\hat{\mathcal{Q}}$ with weight $\theta^-=\theta$.
		\For{\textit{iteration=1 to I}}
		\State \multiline{With probability $\epsilon$ select a random action $a_t$, otherwise select $a_t=\argmax \mathcal{Q}^*(s_t, a_t; \theta)$.}
		\State \multiline{Perform action $a_t$ and observe reward $r_t$ and next state $s_{t+1}$.}
		\State \multiline{Store transition $(s_t, a_t, r_t, s_{t+1})$ in the replay memory $\mathbf{D}$.}
		\State \multiline{Sample random mini-batch of transitions $(s_j, a_j, r_j, s_{j+1})$ from $\mathbf{D}$.}
		\State $y_j=r_j+\gamma\max_{a_{j+1}}\hat{\mathcal{Q}}(s_{j+1},a_{j+1};\theta^-)$.
		\State \multiline{Perform a gradient descent step on $(y_j-\mathcal{Q}(s_j, a_j; \theta))^2$ with respect to the network parameter $\theta$.}
		\State Every $C$ steps reset $\hat{\mathcal{Q}} = \mathcal{Q}$.
		\EndFor
	\end{algorithmic}
\end{algorithm}
In Algorithm~\ref{deepqlearning}, we provide the detail of the deep Q-learning algorithm. In particular, the learning process consists of multiple training steps, i.e., iterations. At each iteration, similar to the Q-learning algorithm, the deep Q-learning algorithm uses the $\epsilon$-greedy mechanism to choose an action. Specifically, with probability $\epsilon$, the algorithm selects a random action. Otherwise, the action that has the highest Q-value will be selected. In practice, the value of $\epsilon$ can be a constant or can be decayed from a large value (i.e., randomized policy) to a small value (i.e., deterministic policy). After performing the selected action, the algorithm observes the intimidate reward and next state of the system. These observations are then stored in the memory pool $\mathbf{D}$. Next, random experiences from the memory pool will be fed to the deep neural network. Then, the network parameters are updated by minimizing the loss function between the target Q-value and the estimated Q-value as follows:
\begin{equation}
\label{lossfunction}
L_i(\theta_i)=\mathbb{E}_{(s,a,r,s')\sim U(\mathbf{D})}\bigg[ \bigg( r + \gamma\max_{a'}\hat{\mathcal{Q}}(s',a';\theta_i^-) -\mathcal{Q}(s,a;\theta_i)\bigg)^2\bigg],
\end{equation}
where $\gamma$ denotes the discount factor, $\theta_i^-$ and $\theta_i$ are the parameters of the Q-network and the target Q-network, respectively. The loss function is then minimized by performing gradient steps on the following gradient:
\begin{equation}
\label{gradient_loss}
\begin{aligned}
\nabla_{\theta_i}L(\theta_i)=\mathbb{E}_{(s,a,r,s')} \bigg[\bigg(r+\gamma \max_{a'}\hat{\mathcal{Q}}(s',a';\theta_i^-) -\mathcal{Q}(s,a;\theta_i)\nabla_{\theta_i}\mathcal{Q}(s,a;\theta_i)\bigg)\bigg].
\end{aligned}
\end{equation}
\section{Optimal Deception Strategy with Deep Dueling Neural Network}
\label{Sec:deepdueling}
\subsection{Deep Dueling Neural Network Architecture}
\begin{figure}[!]
	\centering
	\includegraphics[scale=0.15]{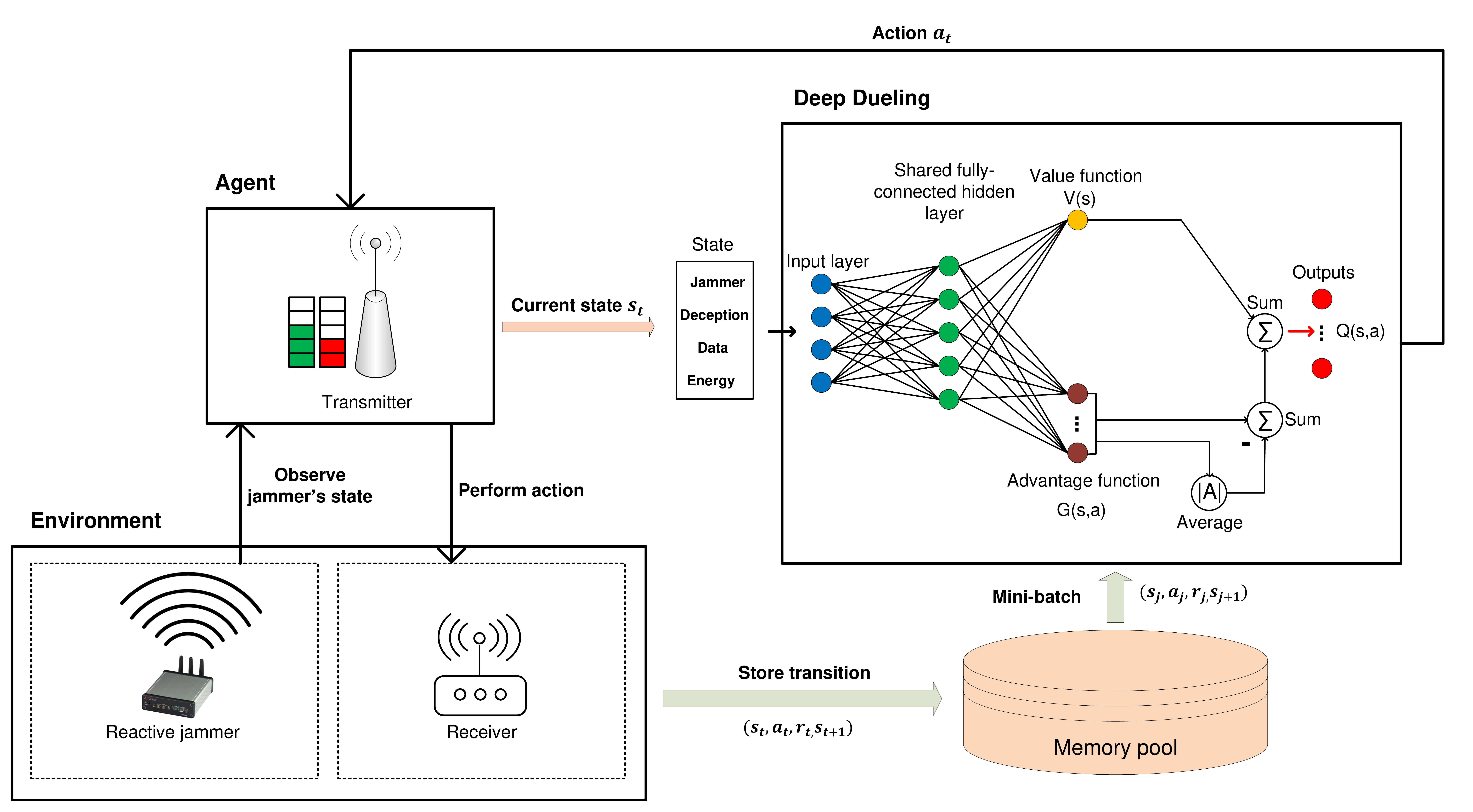}
	\caption{Deep dueling neural network architecture based solution.}
	\label{Fig.deepduelingqlearning}
\end{figure}
\captionsetup{labelfont={color=black}}
In this section, we present the deep dueling algorithm~\cite{Wang2015Dueling} to further improve the convergence rate of the deep Q-learning algorithm, especially when dealing with large state/action spaces as considered in our work. The principle of the deep dueling algorithm is separately estimating the value function and the advantage function. In particular, the value function is used to measure how good it is when the system is at a given state. The advantage function represents the importance of a given action compared to other actions. It can be observed that in many states, performing an action has no effect on the system. Thus, it is unnecessary to estimate the value of this action. For example, performing the rate adaptation technique only matter when the jamming power is low. As a result, unlike the conventional deep Q-learning algorithm, the deep dueling algorithm uses the deep dueling neural network with two streams of hidden layers to estimate the advantage and value functions separately. These two streams are then combined at the output layer to obtain the Q-value for each action at a given state. The proposed deep dueling neural network is illustrated in Fig.~\ref{Fig.deepduelingqlearning}.

It is worth noting that the original deep dueling network~\cite{Wang2015Dueling} was developed for video games, and thus this architecture includes convolutional layers to process images at the input layer. However, in our current work, the input layer consists of the system states, i.e., status of the jammer, deception, data queue and energy queue. As a result, we develop a new deep dueling neural network that contains only fully-connected hidden layers to capture the particular system states of our current work. Note that the deep dueling and deep Q-learning algorithms share the same learning procedure as presented in Section~\ref{sec:QDeepQ}. The difference between these two algorithms is the novel deep dueling neural network architecture. In the following, we discuss on how to construct the deep dueling neural network and the mathematical analysis behind it.

Given policy $\pi$, the Q-value function of state-action pair $(s,a)$ that calculates the value of performing action $a$ at state $s$ can be expressed as follows:
\begin{equation}
	\mathcal{Q}^{\pi}(s,a) = \mathbb{E} \big[r_t|s_t=s, a_{t}=a,\pi\big].
\end{equation}
As such, the value function that determines good it is to be in state $s$ is defined as follows~\cite{Wang2015Dueling}:
\begin{equation}
	\mathcal{V}^{\pi}(s)= \mathbb{E}_{a \sim \pi(s)}\big[\mathcal{Q}^{\pi}(s,a)\big].
\end{equation}
As mentioned, the advantage function is used to measure the importance of each action. Thus, the advantage function of action $a$ can be obtained by decoupling the value function from the Q-function as follows:
\begin{equation}
\mathcal{G}^{\pi}(s,a) =  \mathcal{Q}^{\pi}(s,a) - \mathcal{V}^{\pi}(s).
\end{equation}

The deep dueling algorithm implements two streams of hidden layers to estimate $\mathcal{G}$ and $\mathcal{V}$ function. In particular, one stream procedures a scalar $\mathcal{V}(s;\beta)$ while the other estimates an $|\mathcal{A}|$-dimensional vector $\mathcal{G}(s, a;\alpha)$, where $\alpha$ and $\beta$ are the parameters of the advantage stream and the value stream, respectively. At the output layer, the Q-function is then obtained by combining these two sequences as follows:
\begin{equation}
\label{combined}
\mathcal{Q}(s, a;\alpha, \beta) = \mathcal{V}(s;\beta) + \mathcal{G}(s, a;\alpha).
\end{equation}
Note that $\mathcal{Q}(s, a;\alpha, \beta)$ is a parameterized estimate of the true Q-function. Given $\mathcal{Q}$, it is impossible to derive $\mathcal{V}$ and $\mathcal{G}$ uniquely. This is due to the fact that the Q-value is not changed when subtracting a constant from $\mathcal{G}(s, a;\alpha)$ and adding the same constant to $\mathcal{V}(s;\beta)$. This leads to poor performance as (\ref{combined}) is unidentifiable. To deal with this issue, the Q-value function is obtained by the following mapping:
\begin{equation}
	\label{ouput_max}
	\mathcal{Q}(s,a;\alpha,\beta) = \mathcal{V}(s;\beta) + \big(\mathcal{G}(s,a;\alpha)-\max_{a' \in \mathcal{A}}\mathcal{G}(s,a';\alpha)\big).
\end{equation}
To stable the algorithm, (\ref{ouput_max}) can be converted to a simple form as follows:
\begin{equation}
\label{output_average}
\mathcal{Q}(s,a;\alpha,\beta) = \mathcal{V}(s;\beta) + \big(\mathcal{G}(s,a;\alpha)- \frac{1}{|\mathcal{A}|}\sum_{a'}^{}\mathcal{G}(s, a';\alpha)\big).
\end{equation}
Clearly, in (\ref{output_average}), the algorithm only needs to obtain the average of the advantages instead of finding the max advantage for all the possible actions at state $s$ as in (\ref{ouput_max}). Thus, the algorithm is stabled as demonstrated in~\cite{Wang2015Dueling}.
\begin{algorithm}[!]
	\caption{Deep Dueling Neural Network Based Anti-jamming Algorithm}
	\label{deepduelingqlearning}
	\begin{algorithmic}[1]
		\State Initialize replay memory $\mathbf{D}$ to capacity $\mathcal{D}$.
		\State Initialize the $\mathcal{Q}$ network with random weights $\alpha$ and $\beta$.
		\State Initialize the target $\hat{\mathcal{Q}}$ network with weights $\alpha^- = \alpha$ and $\beta^- = \beta$.
		\For{\textit{iteration=1 to I}}
		\State \multiline{Select action $a_t$ by using the $\epsilon$-greedy mechanism.}
		\State \multiline{Perform action $a_t$ and observe reward $r_t$ and next state $s_{t+1}$.}
		\State Store transition $(s_t, a_t, r_t, s_{t+1})$ in the replay memory.
		\State \multiline{Sample random mini-batch of transitions $(s_j, a_{j}, r_j, s_{j+1})$ from the replay memory.}
		\State \multiline{Combine the value function and advantage functions based on (\ref{output_average}).}
		\State \multiline{$y_j=r_j+\gamma\max_{a_{j+1}}\hat{\mathcal{Q}}(s_{j+1},a_{j+1}; \alpha^-, \beta^-)$.}
		\State \multiline{Perform a gradient descent step on $(y_j-\mathcal{Q}(s_j, a_{j}; \alpha, \beta))^2$.}
		\State Every $C$ steps reset $\hat{\mathcal{Q}} = \mathcal{Q}$.
		\EndFor
	\end{algorithmic}
\end{algorithm}
Through~(\ref{output_average}) and the procedure of the deep Q-learning algorithm, the deep dueling based anti-jamming algorithm is proposed in~Algorithm~\ref{deepduelingqlearning}. It is worth noting that (\ref{output_average}) is deployed as a part of the deep dueling neural network architecture, and thus the value function $\mathcal{V}(s;\beta)$ and the advantage function $\mathcal{G}(s,a;\alpha)$ are estimated without changing the deep Q-learning algorithm. As in this paper, we propose a dynamic MDP framework that can automatically construct two decision epochs in a time slot where the deception mechanism is performed. Thus, at the beginning of each iteration, the algorithm check if the deception mechanism is performed or not to decide the number of decision epochs in the current time slot.
\subsection{Complexity Analysis and Implementation}
\paragraph{Dynamic MDP}
Although our proposed MDP framework will slightly increase the number of states and decision epochs, the performance of the proposed MDP is much better than that of the conventional MDP because the transmitter can deploy the deception mechanism to lure the jammer and leverage the jamming signals. Specifically, for the conventional MDP, the system state space is constructed by the states of the reactive jammer, the data queue, and the energy queue. Thus, the total number of states is $2 \times (D+1) \times (E+1)$. Meanwhile, in our proposed two-period MDP, the system state space is constructed by the state of the deception, the jammer, the data queue, and the energy queue. Note that, the deception strategy has only two states, i.e., deception or not deception. Hence, the total number of states in our proposed MDP is $2 \times 2 \times (D+1) \times (E+1)$. Clearly, the number of states in our proposed MDP is just two times higher than the number of states in the conventional MDP, but it can capture a very important strategy to deal with reactive jammers, i.e., deception strategy. Thus, the performance of the proposed MDP can significantly outperform that of the conventional MDP. Similarly, in our proposed MDP, if the transmitter chooses to perform the deception at the beginning of a time slot, there are two decision epochs in this time slot as discussed in Section III. However, if the transmitter chooses to actively transmit at the beginning of a time slot, there is only one decision epoch in this time slot. Thus, the increase in the number of decision epochs is not significant, i.e., less than two times compared with that of the conventional MDP. In addition, although the number of decision epochs increases, the total number of time slots is the same as that of the conventional MDP, but this increase gives more chances for the transmitter to learn from the actions of the jammer. As a result, our proposed MDP can achieve much better performance compared to the conventional MDP, as discussed in Section~\ref{sec:evaluation}.
\paragraph{Complexity of Training Process}
In the deep dueling neural network used in this work, we implement one input layer $L_0$, one hidden layer $L_1$, and two layers $L_\mathrm{value}$ and $L_\mathrm{advantage}$ to estimate the value and the advantage function, respectively. Denote $|L_\mathrm{i}|$ as the size (i.e., the number of neurons) of layer $L_\mathrm{i}$. The complexity of the deep dueling neural network is $|L_0||L_1| + |L_1||L_\mathrm{value}| + |L_1||L_\mathrm{advantage}|$. In each iteration, a training batch will be fed into the neural network for learning. Denote $N_\mathrm{b}$ as the size of the training batch and $I$ as the total number of training iterations, we have the total complexity of the training process is $\mathcal{O}\Bigg(IN_\mathrm{b}\Big(|L_0||L_1| + |L_1||L_\mathrm{value}| + |L_1||L_\mathrm{advantage}|\Big)\Bigg)$.
\paragraph{Implementation}
In general, the deep neural network requires high computing resources, especially in large-scale system that requires complex network architectures to achieve high accuracy. To make deep learning feasible for resource-constrained devices, e.g., IoT devices, several solutions have been introduced to reduce the complexity of the deep neural network while maintaining a good accuracy. For instance, in~\cite{Han2015Learning}, a compression mechanism is proposed to convert complex neural networks to sparsely connected networks. Through experiments, the authors demonstrated that the computational and storage complexities can be reduced by a factor of 10. Note that our proposed deep dueling neural network is simple with only one fully-connected layer $L_1$. Thus, together with recent advance in network compression and hardware acceleration, the proposed framework can be implemented for general wireless systems.

For ultra-low power IoT devices which cannot implement deep reinforcement learning algorithms, the complex tasks can be offloaded to a nearby resourceful device, e.g., an IoT gateway or a nearby edge computing node. In particular, with current optimal policy stored at the transmitter, the transmitter performs action $a_j$ given state $s_j$. After that, it observes immediate reward $r_j$ and next state $s_{j+1}$. These observations are then stored in the memory of the transmitter. After a certain time, e.g., one hour, the transmitter sends the experiences in its memory to the gateway for training. The optimal policy after training will be sent back to the transmitter to update its current policy. This mechanism is very beneficial in scenarios with multiple transmitters in which the gateway can collect more experiences and improve the learning efficiency.
\section{Performance Evaluation}
\label{sec:evaluation}
\subsection{Parameter Setting}
In the system under consideration, we set the energy storage capacity at $10$ units. The data queue of the transmitter can store up to $10$ packets. The set of jamming power levels is set at $\mathbf{P}_{\mathrm{J}}$ = \{$0$W, $4$W, $10$W, $15$W\}, with $P_{\max}$= $15$W~\cite{15WJammer}. In this work, once detecting activities of the transmitter on the target channel, the reactive jammer can attack the legitimate system at various power levels with different attack probabilities as long as the condition in~(\ref{eq:attack_strategy}) is satisfied. Thus, we set the attack strategy of the jammer to be $\mathbf{J}_s = \{x_0, x_1, x_2, x_3\} = \Big\{1-\frac{P_\mathrm{avg}}{P^\dagger}, 0.5\frac{P_\mathrm{avg}}{P^\dagger}, 0.3\frac{P_\mathrm{avg}}{P^\dagger}, 0.2\frac{P_\mathrm{avg}}{P^\dagger}\Big\}$, where $P^\dagger$ is the maximum average power of the jammer which depends on the hardware configurations and power budget of the jammer. In the simulation, we set $P^\dagger = 10$W. $P_\mathrm{avg}$ is the average power constraint of the jammer, which depends on the jammer's attack strategies. $P_\mathrm{avg}$ will be varied to evaluate the performance of our solution under different jamming strategies. Clearly, with a higher average power constrain $P_\mathrm{avg}$, the probability that the jammer attacks the channel, i.e., $\frac{P_\mathrm{avg}}{P^\dagger}$, is higher. The ratios of attacking with power levels $P^{\mathrm{J}}_1$, $P^{\mathrm{J}}_2$, and $P^{\mathrm{J}}_3$ are set at 0.5, 0.3, and 0.2, respectively. Unless otherwise states, $P_\mathrm{avg}$ is set at $8$W. If the transmitter performs deception and the jammer attacks the channel, the transmitter can harvest energy from or backscatter data through the strong jamming signals. Recall that when the jammer attacks the channel with higher power levels, the transmitter can harvest more energy from the jamming signals and backscatter more packets through the jamming signals. Hence, we set $\mathbf{e}=\{0,2,3,4\}$ and $\widehat{\mathbf{d}}=\{0,1,2,3\}$. For the rate adaptation technique, we set $d^{\mathrm{r}}_m=$\{2, 1, 0\} corresponding to the jamming power levels $\mathbf{P}_{\mathrm{J}}$ = \{$4$W, $10$W, $15$W\}. It is worth noting that $\widehat{\mathbf{d}}$ and $d^{\mathrm{r}}_m$ are the number of packets successfully received at the receiver. Therefore, they already encountered for the channel conditions, e.g., fading, noise, and SNR. Note that our proposed framework does not require the information about the jammer and the environment, e.g., data arrival rate, miss detection probability, and backscatter rate, in advance and can work with any channel models. These parameters as well as the channel model will be learned by the algorithm to obtain the optimal defense policy for the transmitter.
\begin{table}[h]
	\centering
	\caption{\footnotesize PARAMETER SETTING} \label{tabpara} 
	\begin{tabular}{|l|l|l|l|l|l|l|l|l|l|l|}
		\hline
		Symbol& $e_v$ & $\widehat{d}_\mathrm{a}$ & $e_\mathrm{f}$ & $\widehat{d}_\mathrm{de}$ & $e_\mathrm{r}$ & $p_\mathrm{e}$ & $\lambda$  & $K$ & $p_\mathrm{miss}$ \\ \hline
		Value& 1 & 4 & 1 & 3 & 1 & 0.5 &  0.7 & 3 & 0.01   \\ \hline
	\end{tabular}
\end{table}

In the conventional deep neural network of the deep Q-learning algorithm, two fully-connected hidden layers are implemented. For the deep dueling algorithm, value layer $L_\mathrm{value}$ and advantage layer $L_\mathrm{advantage}$ are deployed to separately estimate the value function and the advantage function, respectively. These two layers are connected to a shared hidden layer (after the input layer) and combined at the output layer. The hidden layer consists of $64$ neurons. The activation function of the hidden layer is the tanh function~\cite{Goodfellow2016Deep}. The mini-batch size is $64$. The capacity of the memory pool is $10,000$. For both the deep Q-learning and the deep dueling algorithm, the target Q-network is updated after every $1,000$ training steps. In the $\epsilon$-greedy method, $\epsilon$ is gradually decayed from $1$ to $0.1$. The learning rate and the discount factor are set at 0.001 and 0.95, respectively. In this work, we develop a simulator to simulate the environment using Python, and we use TensorFlow~\cite{TensorFlow} to implement the deep dueling and deep Q-learning algorithms.

To evaluate the proposed solution, we compare its performance with four other schemes: (i) \textit{harvest-then-transmit (HTT)}, (ii) \textit{backscatter mode (BM)}, (iii) \textit{rate adaptation (RA)}, and (iv) \textit{without deception (WD)}.
\begin{itemize}
	\item \textit{HTT:} For this policy, after performing the deception mechanism, the transmitter can harvest energy from the jamming signals or perform the rate adaptation technique if the jammer attacks the channel. The action space of the HTT policy is then defined as $\mathcal{A}_\mathrm{HTT} \triangleq \{a: a \in \{0,1,2,3,4+m\}\}$. This scheme is to evaluate the system performance without using the ambient backscatter technology.
	
	\item \textit{BM:} In this scheme, after performing the deception, the transmitter can use the ambient backscatter technique to transmit data or perform the rate adaptation technique when the jammer attacks the channel. The action space of the BM policy is then defined as $\mathcal{A}_\mathrm{BM} \triangleq \{a: a \in \{0,1,2,4,4+m\}\}$. This policy is to evaluate the system performance without using the RF energy harvesting technique.
	
	\item \textit{RA:} With this policy, after performing the deception, the transmitter can only perform the rate adaptation technique to transmit data if the jammer attacks the channel. The action space of the RA policy is then defined as $\mathcal{A}_\mathrm{RA} \triangleq \{a: a \in \{0,1,2,4+m\}\}$. This scheme is adopted to evaluate the system performance under jamming attacks when the transmitter does not leverage the strong jamming signals.
	
	\item \textit{WD:} With this policy, the transmitter will transmit data as long as it has data and sufficient energy. This scheme is used to show the performance of the system without using our proposed deception strategy.
\end{itemize}
For fair comparisons, the optimal defense policies of the \textit{HTT}, \textit{BM}, and \textit{RA} schemes are derived by the proposed deep dueling algorithm presented in Section~\ref{Sec:deepdueling}. The performance metrics used for evaluation are the average throughput, packet loss, and packet delivery ratio (PDR). Specifically, the average throughout is defined by total number of packets received by the receiver in a time unit. The packet loss corresponds to the average number of dropped packets in each time unit due to the miss detection, jamming attacks, and limit storage of the data queue. Finally, the PDR is defined by the ratio between the total number of packets arrived at the system and the total number of packets successfully transmitted to the receiver.

Our simulation contains three main components, including an agent (i.e., the transmitter), an environment module, and a deep dueling module as illustrated in Fig.~\ref{Fig.deepduelingqlearning}. At each time slot, the transmitter observes the system state from the environment module. The current system state is then fed into the deep dueling model to obtain the corresponding action (based on the $\epsilon$-greedy policy or based on the current optimal policy learned by the deep dueling algorithm). The action will be then sent to the environment module to obtain the immediate reward. As mentioned, the immediate reward is the number of packets successfully received at the receiver which already accounted for the channel model. After receiving the immediate reward from the environment module, the current experience, i.e., current state, action, next state, and immediate reward, is sent to the memory pool of the deep dueling module. Then, at each time slot, the deep dueling algorithm randomly takes a number of experiences from the memory pool and trains the deep dueling neural network to obtain the optimal policy for the transmitter.

\subsection{Simulation Results}
\label{sec:evaluationB}
\subsubsection{Convergence of Deep Reinforcement Learning Algorithms}
\begin{figure}[!]
	\centering
	\includegraphics[scale=0.35]{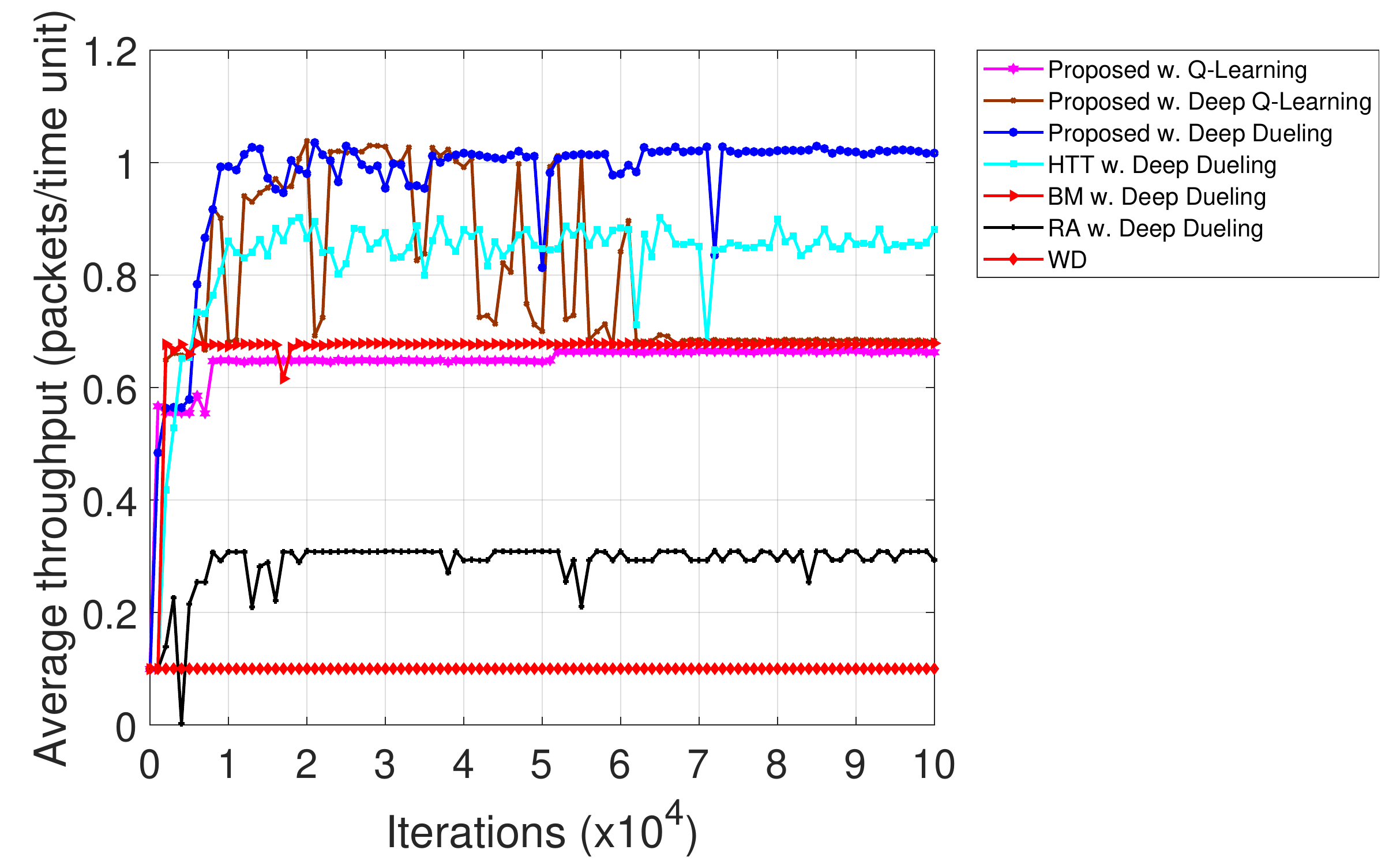}
	\caption{Convergence rates.}
	\label{fig:convergence10}
\end{figure}
Fig.~\ref{fig:convergence10} shows the convergences of our proposed solution together with other baseline solutions using the Q-learning, deep Q-learning, and deep dueling algorithms. It can be observed that, the convergence rate of the Q-learning algorithm is much lower than those of the deep Q-learning and deep dueling algorithms due to the curse-of-dimensionality problem. By using the novel deep dueling neural network architecture, the deep dueling can obtain a very high throughput compared with those of the Q-learning and deep Q-learning algorithms within only $4 \times 10^4$ iterations. For the deep Q-learning algorithm, it cannot obtain the optimal policy within $10^5$ iterations due to the overestimation of the optimizer. In contrast, by using the two separated streams to estimate the advantage and value functions, our proposed deep dueling-based algorithm can converge much faster. Note that the reactive jammer can adjust its attack policy by sensing the activities on the target channel. Thus, with the proposed algorithm, the transmitter can efficiently adapt the optimal defense strategy when the jammer adjusts its policy.

Clearly, the WD policy does not need to learn, but its performance is the worst among all policies. The HTT, BM, and RA policies share a similar convergence time as they all use the proposed deep dueling-based algorithm to obtain their optimal policies. Note that, the BM and RA policies can achieve their optimal policies a bit faster than those of the proposed solution and the HTT policy. The reason is that the proposed solution and the HTT policy can perform the HTT protocol, i.e., harvesting energy when the jammer attacks the channel and uses the harvested energy to actively transmit data when the jammer is idle. As such, the benefit of harvesting energy from the jamming signals will be learned through the immediate reward of using the harvested energy to actively transmit data when the jammer is idle. This makes the learning task more complicated for the proposed solution and the HTT policy compared to those of the RA and BM policies. In the next section, we use the results obtained by the deep dueling algorithm at $4 \times 10^4$ iterations and by the Q-learning algorithm at $10^6$ iterations. Here, the Q-learning algorithm is used as a benchmark to compare with the proposed deep dueling-based algorithm.
\subsubsection{System Performance}
\begin{figure}[!]
	\centering
	\includegraphics[scale=0.35]{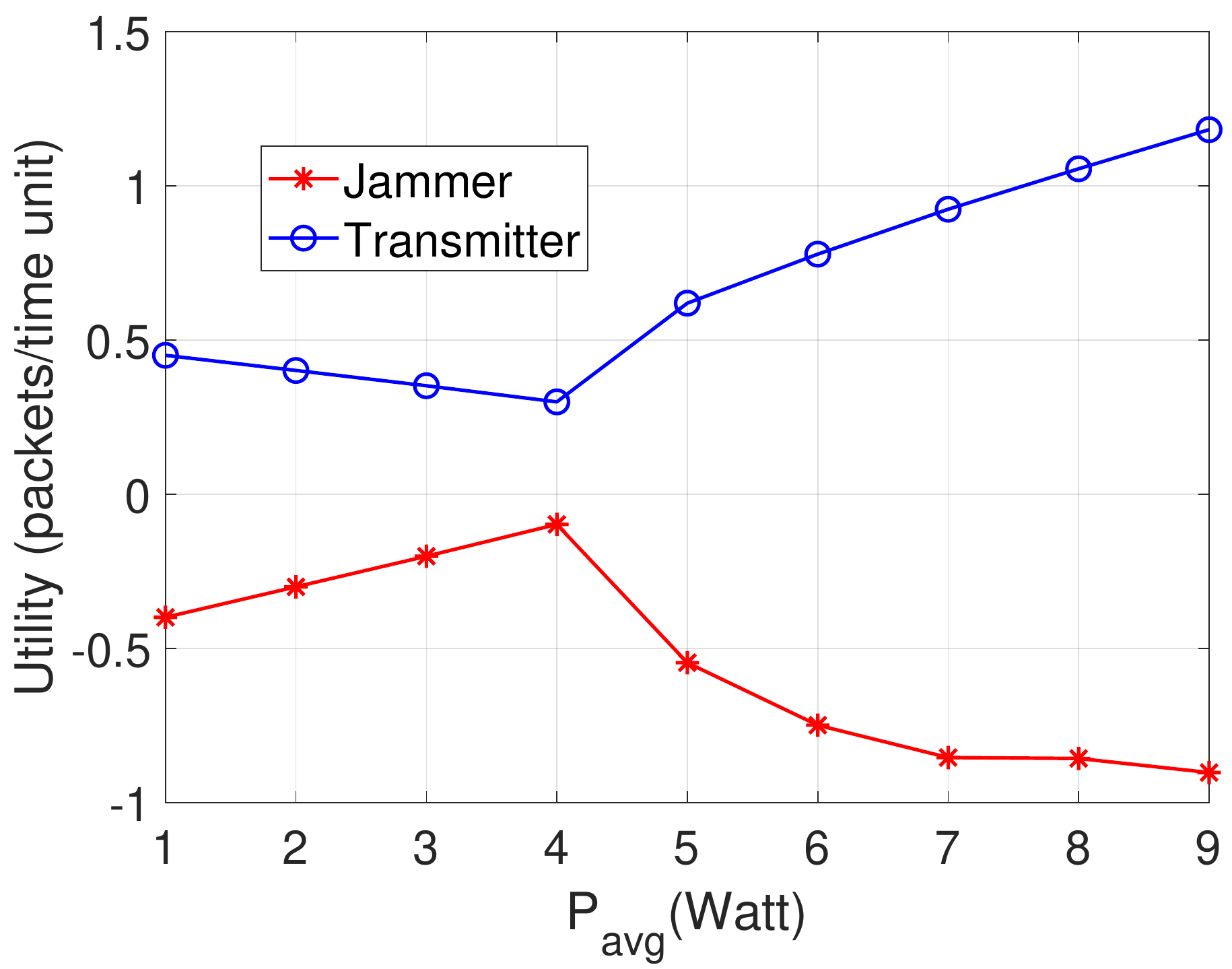}
	\caption{Utilities of the jammer and the transmitter vs. $P_{avg}$.}
	\label{fig:jammer_utility}
\end{figure}
First, we perform simulations to evaluate and compare the utility of the transmitter (under the proposed solution, i.e., Algorithm~\ref{deepduelingqlearning}) with that of the jammer. Specifically, the utility of the transmitter is defined by the average number of packets successfully transmitted to the receiver in a time unit, i.e., the average throughput. The utility of the jammer is defined as in Section~\ref{sec:jammeranalys}. It can be observed that, when the average attack power of the jammer increases from $1$W to $4$W, the average throughput of the proposed solution slightly decreases because the transmitter has fewer opportunities to actively transmit data due to the lack of harvested energy. In addition, harvesting energy from and backscattering data through the jamming signals are not really beneficial when the average attack power of the jammer is low, i.e., less than $4$W. When the power budget of the jammer is larger, i.e., $P_{avg} \geq 4$W, the average throughput obtained by our proposed solution increases as the transmitter has more opportunities to leverage the strong jamming signals to support its transmissions. Thus, the utility of the jammer quickly decreases. This reveals an interesting result that our proposed solution is very effective to deal with reactive jamming attacks even if the jammer has a very high power budget.

\begin{figure*}[!]
	\centering
	\begin{subfigure}[b]{0.3\textwidth}
		\centering
		\includegraphics[scale=0.27]{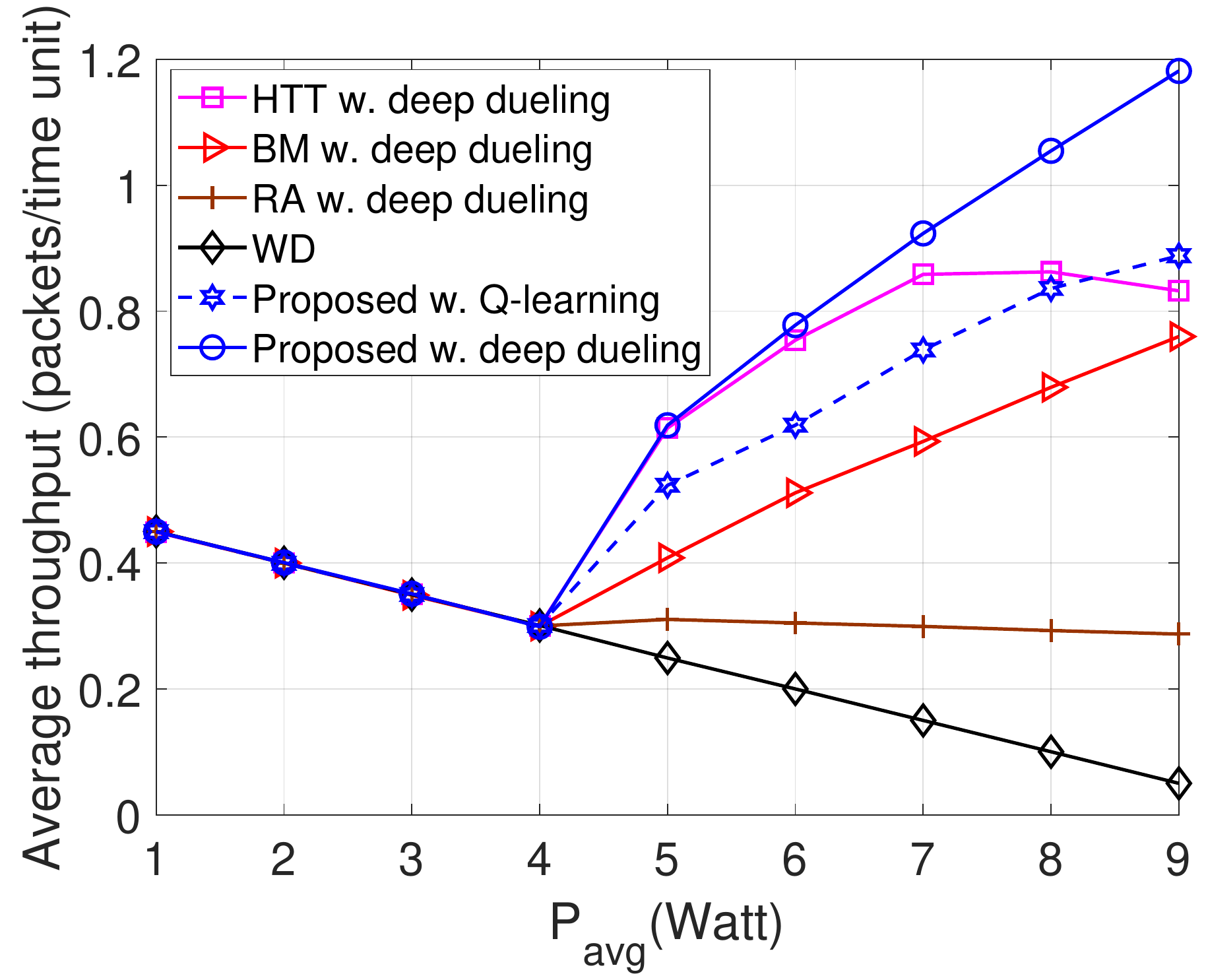}
		\caption{}
	\end{subfigure}%
	~
	\begin{subfigure}[b]{0.3\textwidth}
		\centering
		\includegraphics[scale=0.27]{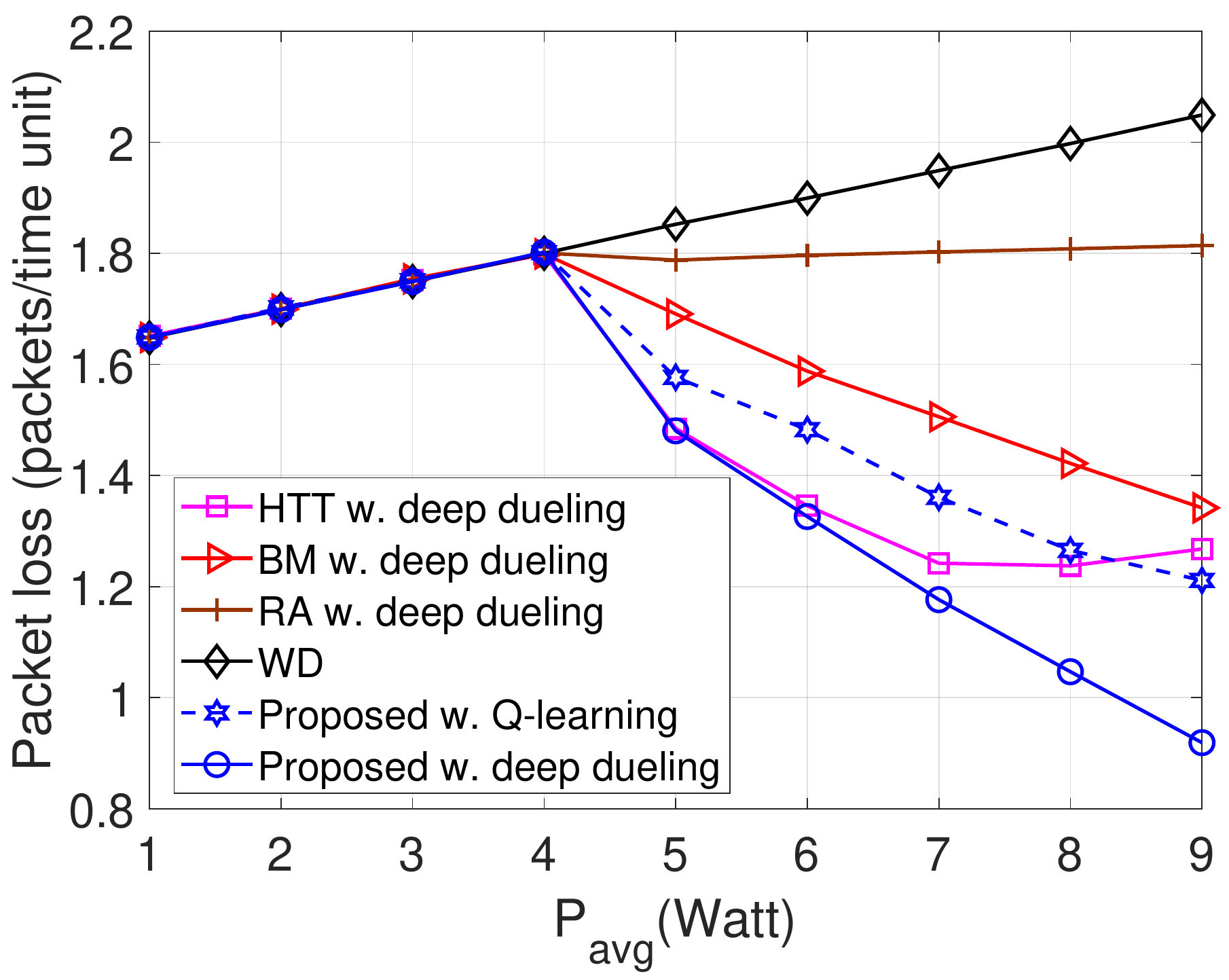}
		\caption{}
	\end{subfigure}%
	~
	\begin{subfigure}[b]{0.3\textwidth}
		\centering
		\includegraphics[scale=0.27]{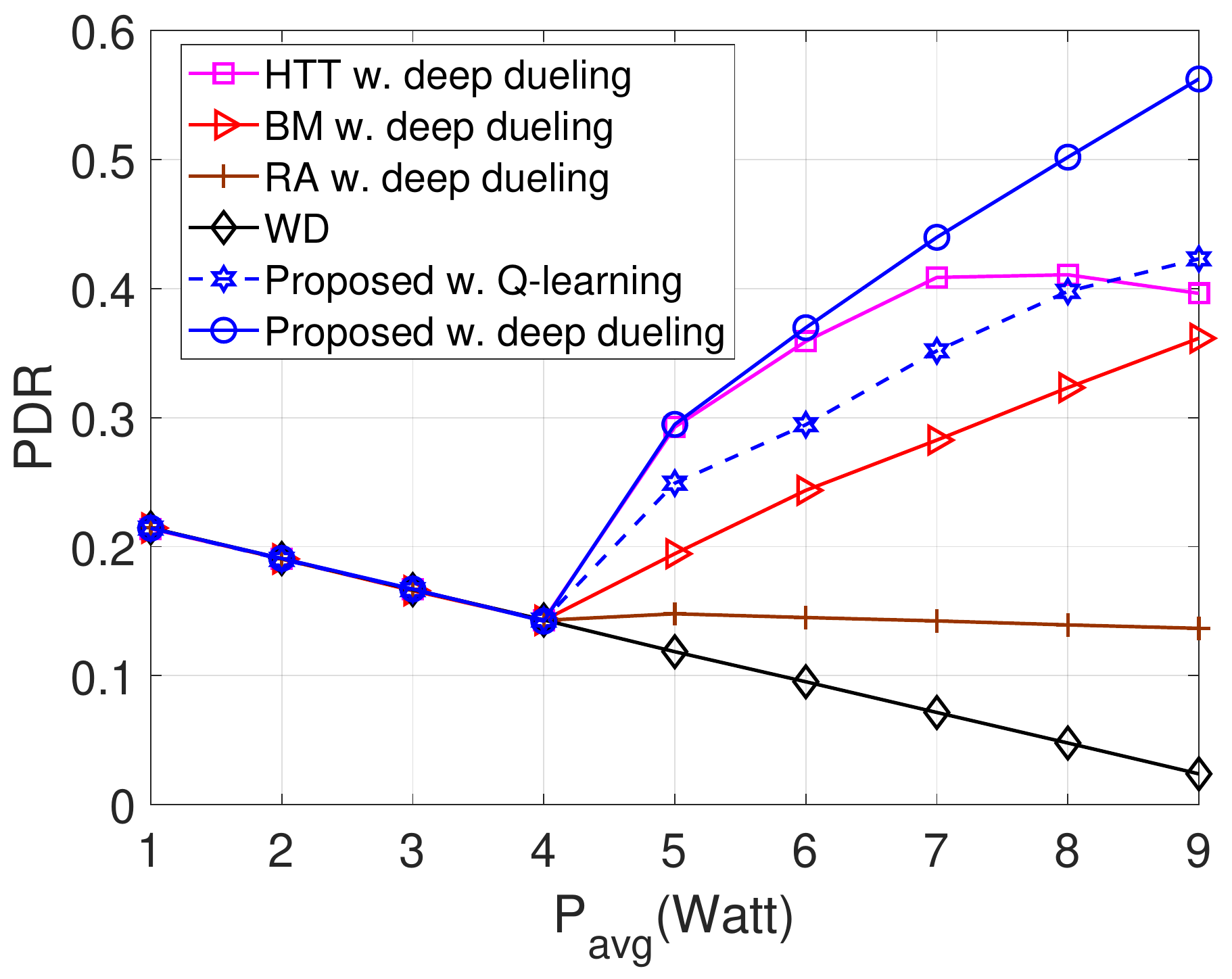}
		\caption{}
	\end{subfigure}%
	\caption{(a) Average throughput (packets/time unit), (b) Packet loss (packets/time unit), and (c) PDR vs. $P_{avg}$.} 
	\label{fig:varyAttack}
\end{figure*}

In Fig.~\ref{fig:varyAttack}, we vary the average attack power $P_{avg}$ of the jammer and observe the performance of the system under different strategies. The performance of the proposed solution can be explained as above. It can be observed that for the BM, HTT and RA strategies, the transmitter only can backscatter data, harvest energy, or reduce the data rate, respectively, under the jamming attack, and thus their performances are much lower than that of the proposed solution which can optimize and trade-off all activities. Note that for the \textit{HTT} policy, the average throughput increases when $P_{avg}$ increases from $4$W to $7$W and decreases when the average attack power of the jammer is high. This is due to the fact that when the jammer often attacks the channel, the transmitter has fewer opportunities to actively transmit data to the receiver. For the \textit{WD} policy, the average throughput decreases when the attack probability increases. The reason is that under this policy, the transmitter can only use the harvested energy from the surrounding environment to actively transmit data when the jammer does not attack the channel. In Fig.~\ref{fig:varyAttack}(b), we show the average number of packet loss of the system. Obviously, the packet loss obtained by our proposed solution is always much lower than those of the other solutions. As a result, the PDR obtained by our proposed solution is higher than those of other solutions as shown in Fig.~\ref{fig:varyAttack}(c).

\begin{figure*}[!]
	\centering
	\begin{subfigure}[b]{0.3\textwidth}
		\centering
		\includegraphics[scale=0.27]{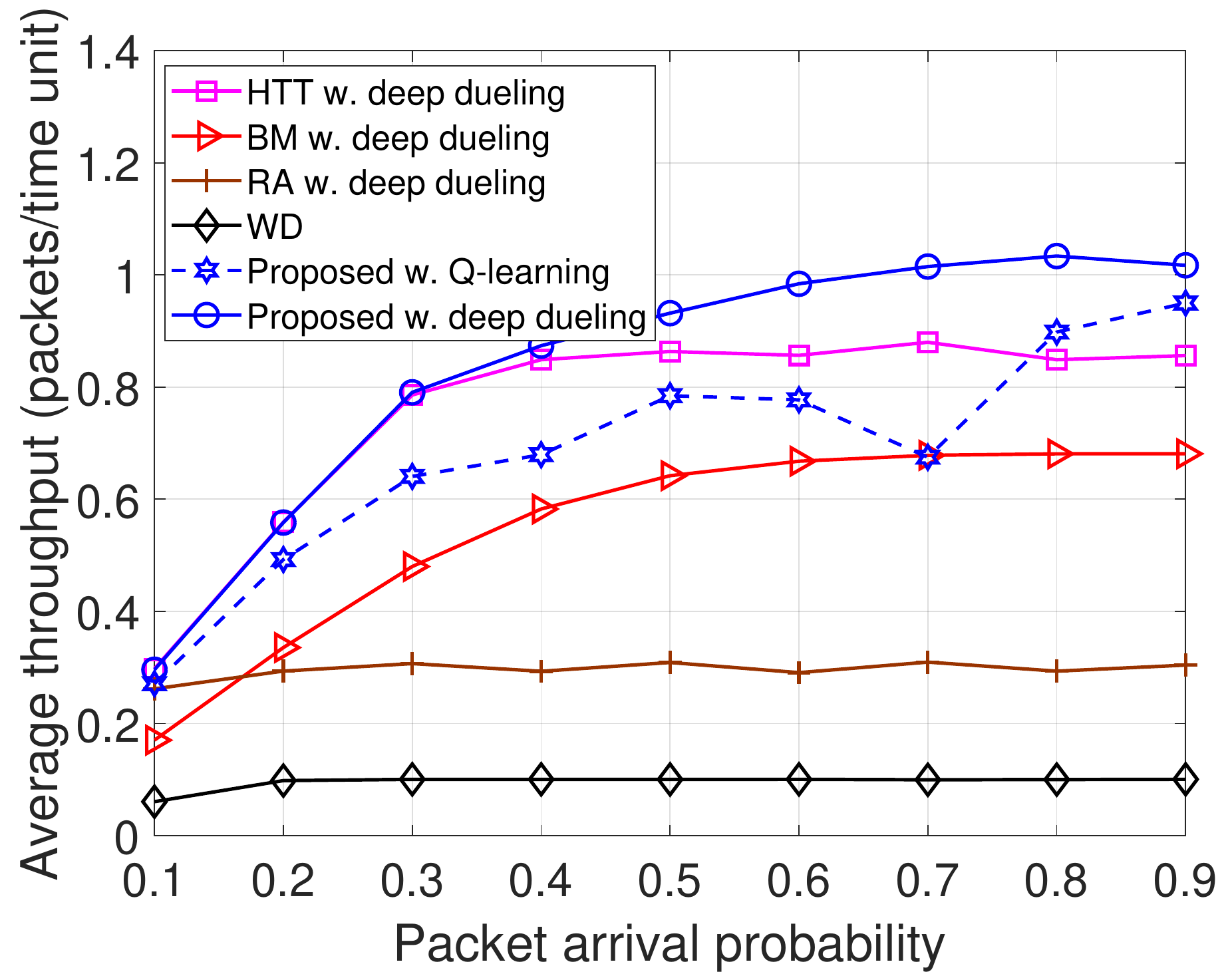}
		\caption{}
	\end{subfigure}%
	~
	\begin{subfigure}[b]{0.3\textwidth}
		\centering
		\includegraphics[scale=0.27]{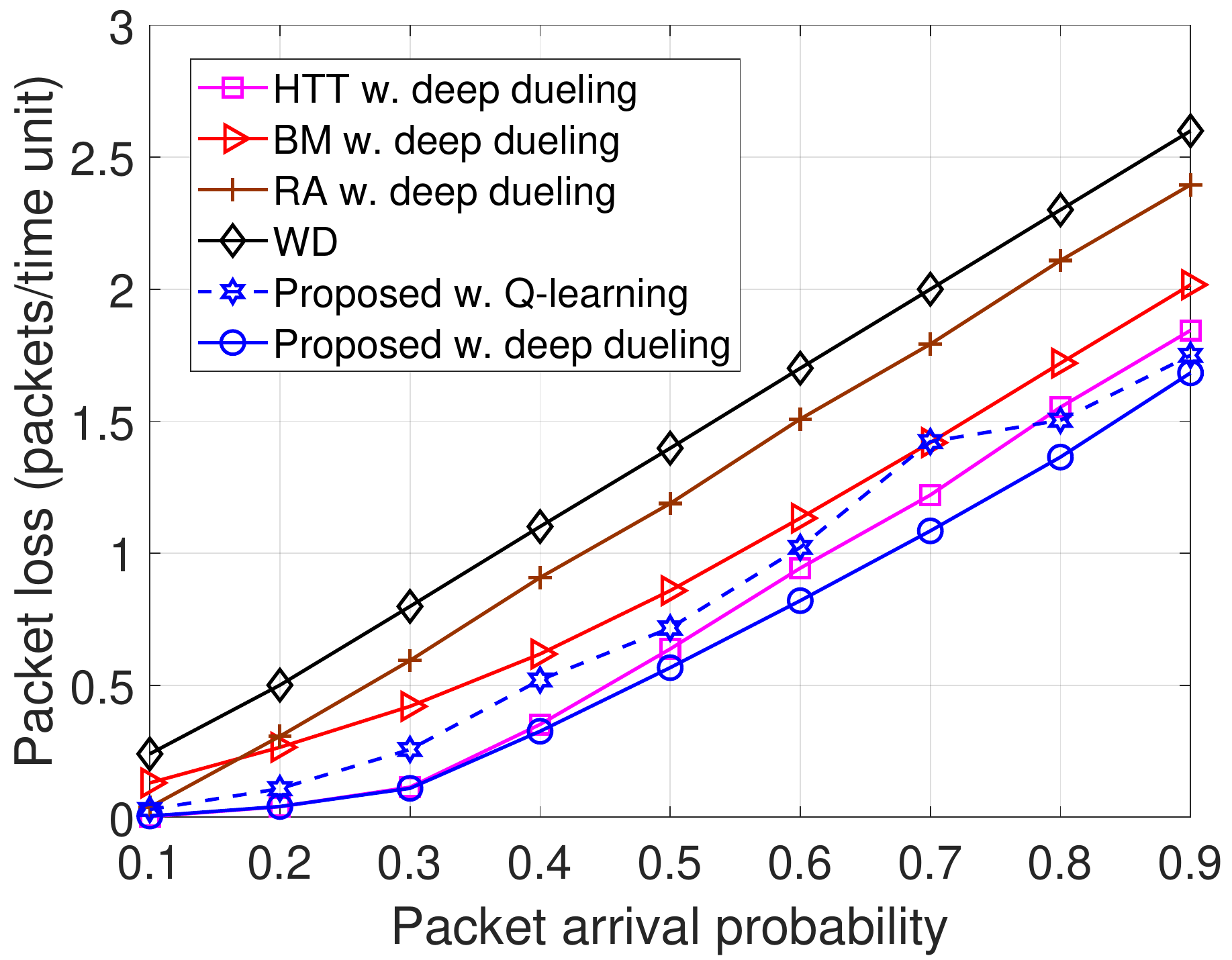}
		\caption{}
	\end{subfigure}%
	~
	\begin{subfigure}[b]{0.3\textwidth}
		\centering
		\includegraphics[scale=0.27]{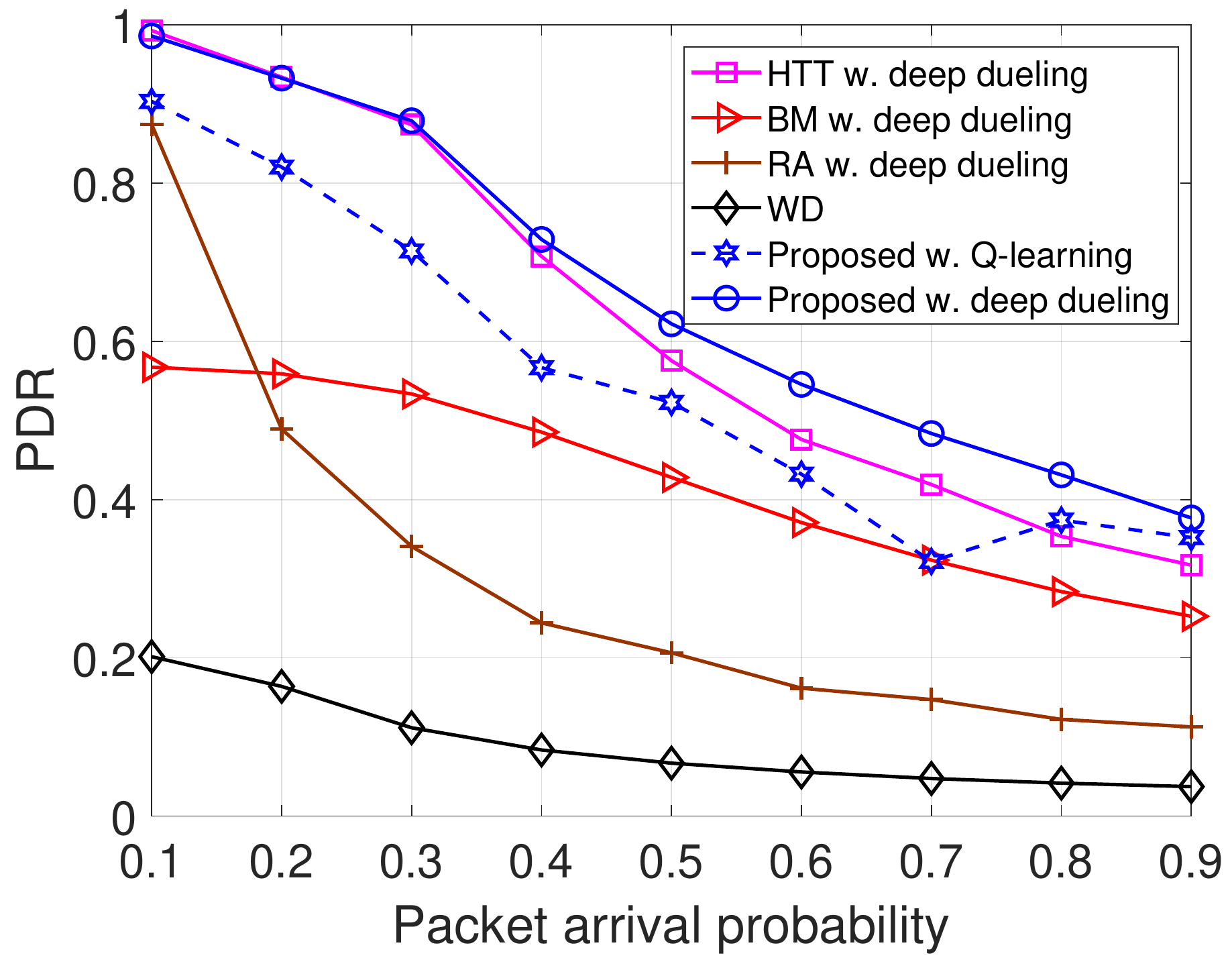}
		\caption{}
	\end{subfigure}%
	
	\caption{(a) Average throughput (packets/time unit), (b) Packet loss (packets/time unit), and (c) PDR vs. packet arrival probability.} 
	\label{fig:varyArrival}
\end{figure*}

Next, in Fig.~\ref{fig:varyArrival}, we vary the packet arrival probability and evaluate the system performance in terms of average throughput, packet loss, and PDR under different policies. As shown in Fig.~\ref{fig:varyArrival}(a), when $\lambda$ increases from $0.1$ to $0.6$, the average throughput obtained by the proposed deep dueling algorithm increases as the transmitter can transmit more packets. Nevertheless, when $\lambda > 0.6$, the average throughput remains stable because the system reaches to the saturation state. Note that the proposed algorithm can always achieve the highest throughput. As such, the PDR obtained by the proposed solution is also higher than those of other schemes as shown in Fig.~\ref{fig:varyArrival}(b). Note that the PDR decreases with $\lambda$ as the total number of packets arrived in the system increases. Moreover, as shown in Fig.~\ref{fig:varyArrival}(b), the packet loss of the system increases proportionally to the packet arrival rate as the data queue size is limited and the transmitter cannot send all arrival packets to the receiver. Note that the performance of the Q-learning algorithm is not as high as the deep dueling algorithm because it cannot obtain the optimal defense policy after $10^6$ training steps.

\begin{figure*}[!]
	\centering
	\begin{subfigure}[b]{0.3\textwidth}
		\centering
		\includegraphics[scale=0.27]{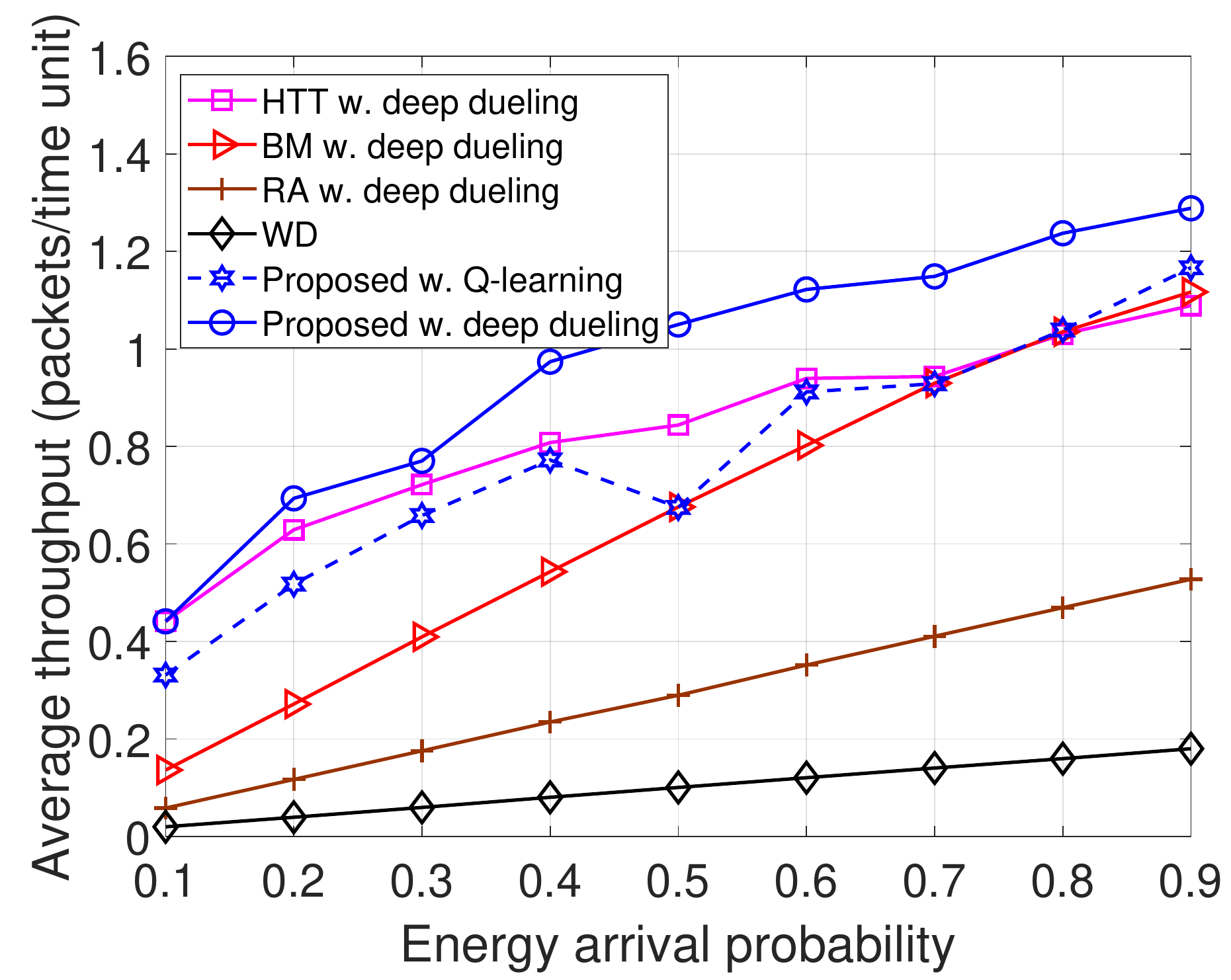}
		\caption{}
	\end{subfigure}%
	~
	\begin{subfigure}[b]{0.3\textwidth}
		\centering
		\includegraphics[scale=0.27]{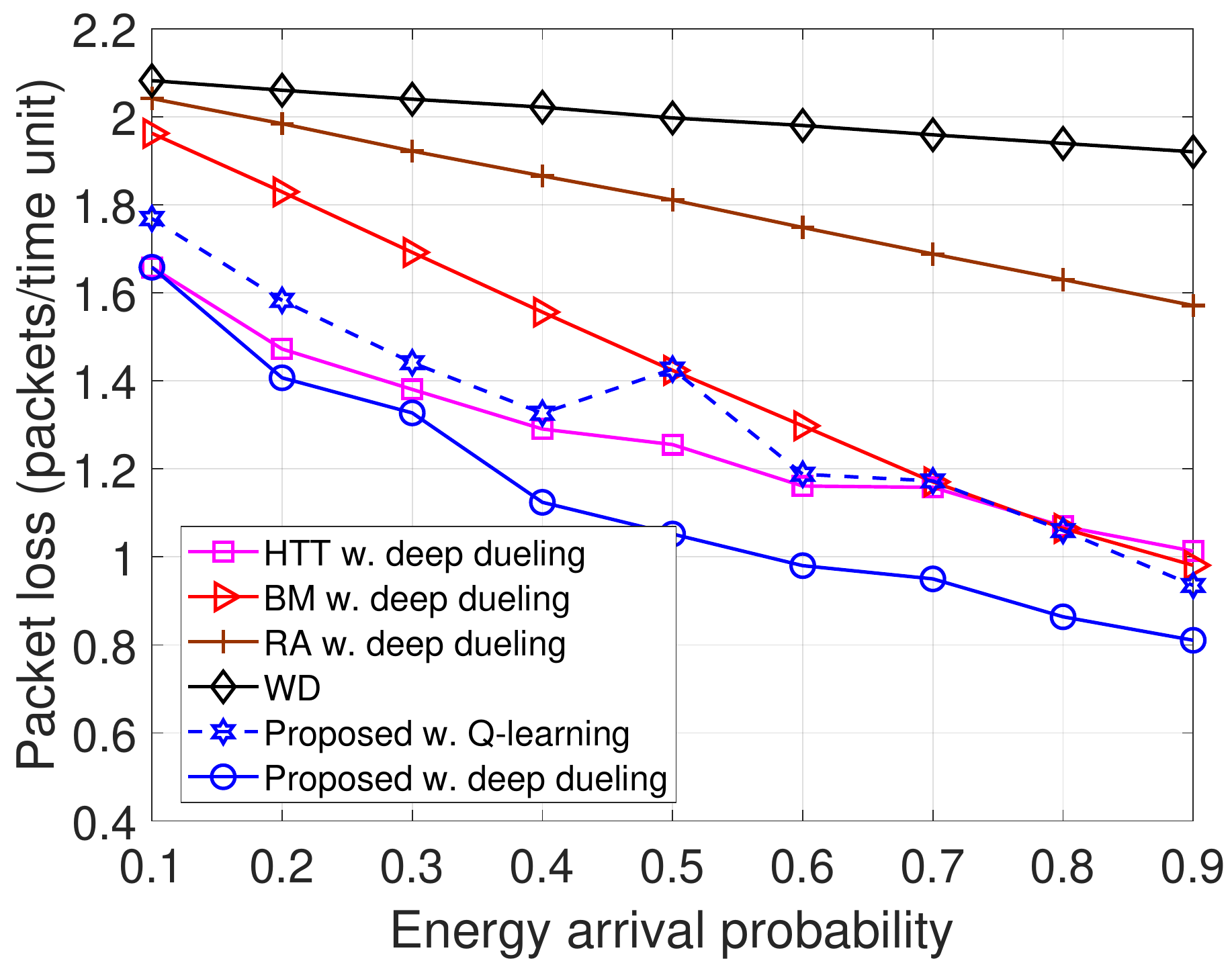}
		\caption{}
	\end{subfigure}%
	~
	\begin{subfigure}[b]{0.3\textwidth}
		\centering
		\includegraphics[scale=0.27]{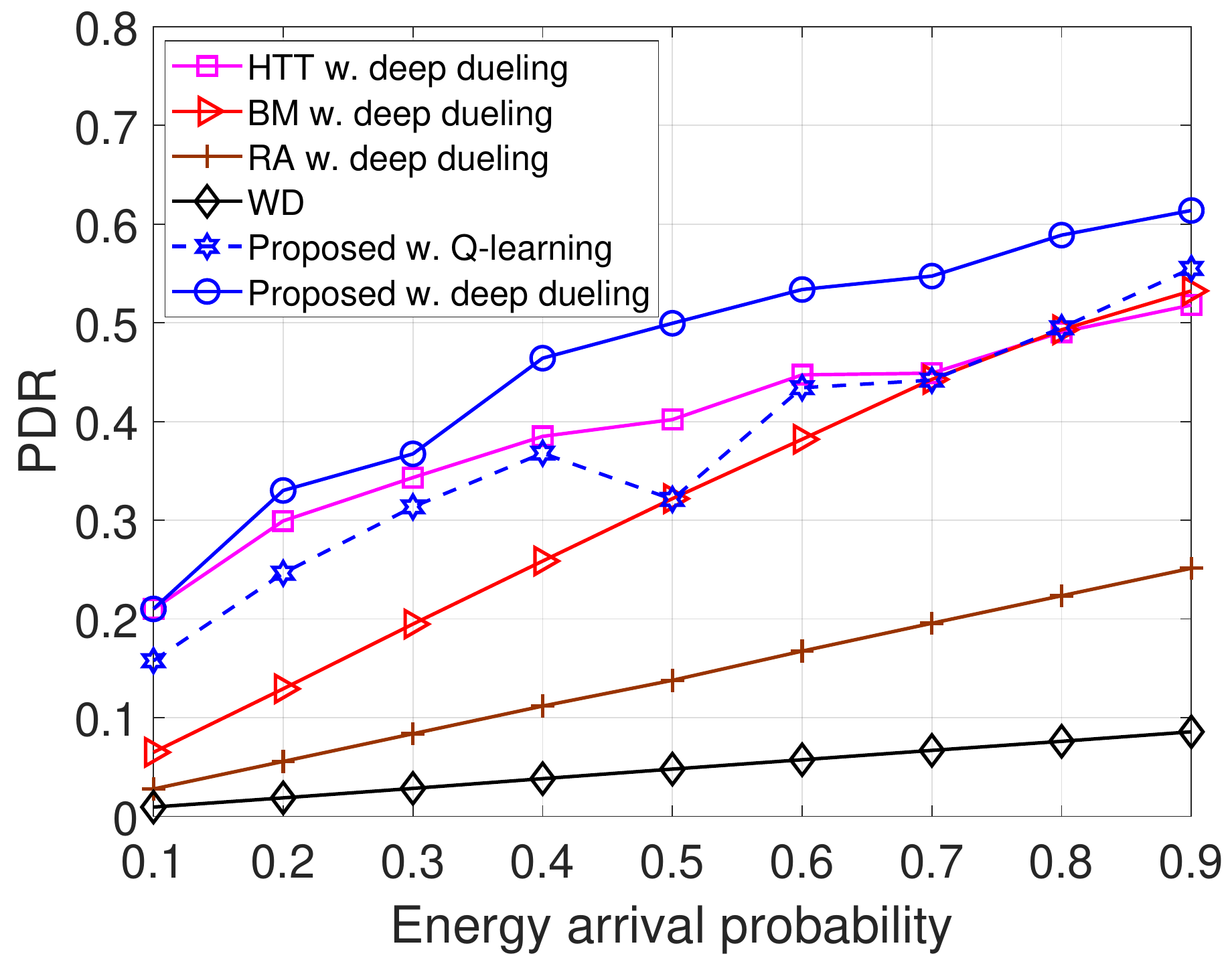}
		\caption{}
	\end{subfigure}%
	\caption{(a) Average throughput (packets/time unit), (b) Packet loss (packets/time unit), and (c) PDR vs. energy arrival probability.} 
	\label{fig:varyEnergyArrival}
\end{figure*}

Finally, we vary the probability that the transmitter can successfully harvest one unit of energy from the environment as shown in Fig.~\ref{fig:varyEnergyArrival}. As shown in Fig.~\ref{fig:varyEnergyArrival}(a), when $p_\mathrm{e}$ increases, the average throughput of the system obtained by all the solutions increases as the transmitter can harvest more energy from the surrounding environment to support its operations, i.e., deception or active transmissions. This leads to the reduction of the packet loss as shown in Fig.~\ref{fig:varyEnergyArrival}(b) and the increase of the PDR in Fig.~\ref{fig:varyEnergyArrival}(c). In overall, by optimizing the time for data backscattering, energy harvesting, and rate adaptation, our proposed solution always achieves the highest throughput and lowest packet loss compared to those of the other solutions. Again, the performance of the Q-learning algorithm is worse than that of the deep dueling algorithm because of the slow-convergence problem.
\section{Conclusion}
\label{sec:conclusion}
In this paper, we have proposed the intelligent anti-jamming framework which allows the transmitter to effectively defeat powerful reactive jamming attacks. Specifically, with the deception mechanism, the transmitter can perform the deception strategy to attract the jammer and drain its power. Furthermore, we introduce the novel ideas of using recent advanced technologies, i.e., ambient backscatter communications and RF energy harvesting, to enable the transmitter to leverage the strong jamming signals while being attacked to further improve system performance. To deal with the dynamic and uncertainty of the jammer and the environment, we have developed the dynamic MDP framework to capture the special properties of the deception mechanism. Then, the deep dueling algorithm is proposed to improve the learning rate of the transmitter. Extensive simulations have demonstrated that the proposed solution can successfully defeat reactive jamming attacks even with very high attack power levels. Interestingly, we have shown that by leveraging the jamming signals, the more frequently the jammer attacks the channel, the greater performance the system can achieve. Moreover, our proposed solution can improve the average throughput by up to 20 times higher compared to the solution without using the deception strategy. One of the potential research directions from this work is to study the interaction between our proposed approach with a reactive jammer empowered by reinforcement learning algorithms. In this case, stochastic game model could be an effective tool to study strategies of the jammer as well as evaluate the system performance.
\appendices
\section{The proof of Theorem~\ref{theo:limitexists}}
\label{appendix:limitexist}
To prove this theorem, we first show that the Markov chain is irreducible. It means that the process can go from one state to any other state after a finite number of steps. In particular, in the system under consideration, the reactive jammer only attacks the channel when the transmitter actively transmits data. Thus, at the beginning of a time slot, the state of the jammer is always idle, i.e., $j=0$. Similarly, the state of the deception always equals $0$ at the beginning of a time slot. As a result, at the beginning of a time slot, the system state is $s=(0,0,d,e)$. If the transmitter chooses to perform the deception and the jammer attacks the channel, the system moves to state $s'=(1,1,d,e-e_\mathrm{f})$. If the jammer does not attack the channel, the system moves to state $s'=(1,0,d,e-e_\mathrm{f})$. Thus, from given states of the jammer and the deception, the system can move to any other states of the jammer and the deception after a finite number of steps.

Similarly, from given data and energy states, the system can move to any other states of the data and energy queues after a finite number of steps. In particular, if the transmitter chooses to actively transmit $d_\mathrm{a}$ packets to the receiver, the the data state moves from $d$ to $d-d_\mathrm{a}$, and the energy state moves from $e$ to $e-(d_\mathrm{a} \times e_\mathrm{r})$. If the transmitter chooses to perform the deception, the energy state will move from $e$ to $e-e_\mathrm{f}$. After performing the deception, if the jammer attacks the channel and the transmitter chooses to harvest energy, the energy state moves from $e$ to $e+e^\mathrm{J}_n$. If the transmitter chooses to backscatter data, the data state moves from $d$ to $d-d^\mathrm{J}_n$. If the transmitter chooses to adapt its rate, the data state moves from $d$ to $d-d^\mathrm{r}_m$ and the energy state moves from $e$ to $e-(d^\mathrm{r}_m \times e_\mathrm{r})$. In contrast, if the jammer does not attack the channel and the transmitter chooses to actively transmit data, the data state moves from $d$ to $d-d_\mathrm{de}$ and the energy state moves from $e$ to $e-(d_\mathrm{de} \times e_\mathrm{r})$. If there are $K$ packets arriving at the system, the data state will move from $d$ to $d+K$. If the transmitter can successfully harvest energy from the ambient RF signals, the energy state moves from $e$ to $e+1$.

Thus, the state space $\mathcal{S}$ (which is a combination of the states of the jammer, the deception, the data queue, and the energy queue) contains only one communicating class, i.e., from a given state the process can move to any other states after a finite number of steps. In other words, the MDP with states in $\mathcal{S}$ is irreducible. As such, the average throughput $\mathcal{R}(\pi)$ is well defined and does not depend on the initial state for every $\pi$~\cite{CompetitiveBook}. Thus, we can always obtain the optimal policy for the transmitter regardless of the initial system state.


\begin{thebibliography}{100}
\bibliographystyle{IEEEtranS}

\bibitem{Hanawal2016Joint}
M.~K.~Hanawal, M.~J.~Abdel-Rahman, and M.~Krunz, ``Joint adaptation of frequency hopping and transmission rate for anti-jamming wireless systems,'' \emph{IEEE Transactions on Mobile Computing}, vol. 15, no. 9, pp. 2247-2259, Sept. 2016.

\bibitem{Mpitziopoulos2009Survey}
A.~Mpitziopoulos, D.~Gavalas, C.~Konstantopoulos, and G.~Pantziou, ``A survey on jamming attacks and countermeasures in WSNs,'' \emph{IEEE Communications Surveys \& Tutorials}, vol. 11, no. 4, pp. 42-56, Fourth Quarter 2009.

\bibitem{Xu2006Jamming}
W.~Xu, K.~Ma, W.~Trappe, and Y.~Zhang, ``Jamming sensor networks: attack and defense strategies,'' \emph{IEEE Network}, vol. 20, no. 3, pp. 41-47, May 2006.

\bibitem{Quan2015Anti}
H.~Quan, H.~Zhao, and P.~Cui, ``Anti-jamming frequency hopping system using multiple hopping patterns,'' \emph{Wireless Personal Communications}, vol. 81, no. 3, pp. 1159-1176, Apr. 2015.

\bibitem{Mpitziopoulos2007Hybrid}
A.~Mpitziopoulos, D.~Gavalas, G.~Pantziou, and C.~Konstantopoulos, ``Defending wireless sensor networks from jamming attacks,'' \emph{IEEE PIMRC}, Athens, Greece, Dec. 2007.

\bibitem{Wang2011Anti}
B.~Wang, Y.~Wu, K.~R.~Liu, and T.~C.~Clancy, ``An anti-jamming stochastic game for cognitive radio networks,'' \emph{IEEE Journal on Selected Areas in Communications}, vol. 29, no. 4, pp. 877-889, Apr. 2011.

\bibitem{Gao2018Game}
Y.~Gao, Y.~Xiao, M.~Wu, M.~Xiao, and J.~Shao, ``Game theory-based anti-jamming strategies for frequency hopping wireless communications,'' \emph{IEEE Trans. Wireless Commun.}, vol. 17, no. 8, pp. 5314-5326, Aug. 2018.

\bibitem{Wu2011Anti}
Y.~Wu, B.~Wang, K.~R.~Liu, and T.~C.~Clancy, ``Anti-jamming games in multi-channel cognitive radio networks,'' \emph{IEEE Journal on Selected Areas in Communications}, vol. 30, no. 1, pp. 4-15, Jan. 2012.

\bibitem{Pelechrinis2009RA}
K.~Pelechrinis, I.~Broustis, S.~V.~Krishnamurthy, and C.~Gkantsidis, ``Ares: An anti-jamming reinforcement system for 802.11 networks,'' \emph{ACM CoNEXT}, Rome, Italy, Dec. 2009.

\bibitem{Firouzbakht2012RA}
K.~Firouzbakht, G.~Noubir, and M.~Salehi,``On the capacity of rate-adaptive packetized wireless communication links under jamming,'' in \emph{Proceedings of the fifth ACM conference on Security and Privacy in Wireless and Mobile Networks}, Tucson, AZ, USA, 2012, pp. 3-14.

\bibitem{Noubir2011RA}
G.~Noubir, R.~Rajaraman, B.~Sheng, and B.~Thapa, ``On the robustness of IEEE 802.11 rate adaptation algorithms against smart jamming,'' in \emph{Proceedings of the fourth ACM conference on Wireless network security}, Hamburg, Germany, Jun. 2011, pp. 97-108.

\bibitem{Gu02017Exploiting}
J.~Guo, N.~Zhao, F.~R.~Yu, X.~Liu, and V.~CM.~Leung, ``Exploiting adversarial jamming signals for energy harvesting in interference networks,'' \emph{IEEE Transactions on Wireless Communications}, vol. 16, no. 2, pp. 1267-1280, Feb. 2017.

\bibitem{Zhao2017Antijamming}
N.~Zhao, J.~Guo, F.~R.~Yu, M.~Li, and V.~CM.~Leung, ``Antijamming Schemes for Interference-Alignment-Based Wireless Networks,'' \emph{IEEE Transactions on Vehicular Technology}, vol. 66, no. 2, pp. 1271-1283, Feb. 2017.

\bibitem{Guo2019Proactive}
J.~Guo, N.~Zhao, Z.~Yang, F.~R.~Yu, Y.~Chen, and V.~CM.~Leung, ``Proactive Jamming Toward Interference Alignment Networks: Beneficial and Adversarial Aspects,'' \emph{IEEE Systems Journal}, vol. 13, no. 1, pp. 412-423, Mar. 2019.

\bibitem{Huynh2019Jam}
N.~V.~Huynh, D.~N.~Nguyen, D.~T.~Hoang, and E.~Dutkiewicz, ``Jam Me If You Can: Defeating Jammer with Deep Dueling Neural Network Architecture and Ambient Backscattering Augmented Communications'', \emph{IEEE Journal on Selected Areas in Communications}, vol. 37, no. 11, pp. 2603-2620, Nov. 2019.

\bibitem{Liu2013Ambient}
V.~Liu, A.~Parks, V.~Talla, S.~Gollakota, D.~Wetherall, and J.~R.~Smith, ``Ambient backscatter: Wireless communication out of thin air,'' \emph{ACM SIGCOMM}, Hong Kong, China, Aug. 2013.

\bibitem{Hoang2020Borrowing}
D.~T.~Hoang, D.~N.~Nguyen, M.~A.~Alsheikh, S.~Gong, E.~Dutkiewicz, D.~Niyato, and Z.~Han, ````Borrowing Arrows with Thatched Boats'': The Art of Defeating Reactive Jammers in IoT Networks,'' \emph{IEEE Wireless Communications}, vol. 27, no. 3, pp. 79-87, Jun. 2020.

\bibitem{Fang2016Wireless}
S.~Fang, Y.~Liu, and P.~Ning, ``Wireless Communications Under Broadband Reactive Jamming Attacks,'' \emph{IEEE Transactions on Dependable and Secure Computing}, vol. 13, no. 3, pp. 394-408, May/Jun. 2016.

\bibitem{Xu2005The}
W.~Xu, W.~Trappe, Y.~Zhang, and T.~Wood, ``The Feasibility of Launching and Detecting Jamming Attacks in Wireless Networks,'' \emph{ACM International Symposium on Mobile Ad Hoc Networking and Computing}, Urbana-Champaign, IL, USA, May 2005.

\bibitem{Lichtman2012Reinforcement}
M.~Lichtman and J.~H.~Reed, ``Reinforcement Learning for Reactive Jamming Mitigation,'' \emph{Journal of Cyber Security and Mobility}, vol. 3, no. 2, pp. 213-230, Apr. 2014.

\bibitem{Rezgui2019Mitigating}
G.~Rezgui, E.~V.~Belmega, and A.~Chorti, ``Mitigating Jamming Attacks Using Energy Harvesting,'' \emph{IEEE Wireless Communications Letters}, vol. 8, no. 1, pp. 297-300, Feb. 2019.

\bibitem{Guo2019Nocoherent}
H.~Guo, Q.~Zhang, D.~Li, and Y.-C.~Liang, ``Noncoherent Multiantenna Receivers for Cognitive Backscatter System with Multiple RF Sources,'' [Online]. Available: arXiv:1808.04316.

\bibitem{Zhang2019Constellation}
Q.~Zhang, H.~Guo, Y.-C.~Liang, and X.~Yuan, ``Constellation Learning-Based Signal Detection for Ambient Backscatter Communication Systems,'' \emph{IEEE Journal on Selected Areas in Communications}, vol. 37, no. 2, pp. 452-463, Feb. 2019.

\bibitem{Boyer2014Backscatter}
C.~Boyer and S.~Roy, ``Backscatter communication and RFID: Coding, energy, and MIMO analysis, \emph{IEEE Transactions on Communications}, vol. 62, no. 3, pp. 770-785, Mar. 2014.
\bibitem{Kimionis2012Bistatic}
J.~Kimionis, A.~Bletsas, and J.~N.~Sahalos, ``Bistatic backscatter radio for tag read-range extension,'' in \emph{Proceedings of IEEE International Conference on RFID-Technologies and Applications (RFID-TA)}, Nice, France, Nov. 2012, pp. 356-361.

\bibitem{Huynh2018Survey}
N.~V.~Huynh, D.~T.~Hoang, X.~Lu, D.~Niyato, P.~Wang, and D.~I.~Kim, ``Ambient Backscatter Communications: A Contemporary Survey,'' \emph{IEEE Communications Surveys \& Tutorials}, vol. 20, no. 4, pp. 2889-2922, Fourthquarter 2018.

\bibitem{Balanis2012Antenna}
C.~A.~Balanis, \emph{Antenna Theory: Analysis and Design}. New York, NY: Wiley, 2012.

\bibitem{CompetitiveBook}
J.~Filar and K.~Vrieze,	\emph{Competitive Markov Decision Processes}. Springer Press, 1997.	

\bibitem{Watkins1992QLearning}
C.~J.~C.~H.~Watkins and P.~Dayan, ``Q-learning,'' \emph{Mach. Learn.}, vol. 8, no. 3–4, pp. 279–292, 1992.

\bibitem{Mnih2015Human}
V.~Mnih, K.~Kavukcuoglu, D.~Silver, A.~A.~Rusu, J.~Veness, M.~G.~Bellemare, A.~Graves, et al., ``Human-level control through deep reinforcement learning,'' \emph{Nature}, vol. 518, no. 7540, pp. 529-533, Feb. 2015.	

\bibitem{Goodfellow2016Deep}
I.~Goodfellow, Y.~Bengio, and A.~Courville, \emph{Deep learning}. MIT press, 2016.	

\bibitem{Wang2015Dueling}
Z.~Wang, T.~Schaul, M.~Hessel, H.~V.~Hasselt, M.~Lanctot, and N.~D.~Freitas, ``Dueling network architectures for deep reinforcement learning,'' [Online]. Available: arXiv:1511.06581.

\bibitem{Han2015Learning}
S.~Han, J.~Pool, J.~Tran, and W.~Dally, ``Learning both weights and connections for efficient neural network,'' in \emph{Proceedings of the 28th International Conference on Neural Information Processing Systems (NIPS)}, Montreal, Canada, Dec. 2015.

\bibitem{15WJammer}
15W Jammer. [Online]. Available: \url{http://www.jammerall.com/products/Adjustable-6-Antenna-15W-High-Power-WiFi\%2CGPS\%2CMobile-Phone-Jammer.html}

\bibitem{TensorFlow}
M.~Abadi, P.~Barham, J.~Chen, Z.~Chen, A.~Davis, J.~Dean, M.~Devin et al, ``TensorFlow: A System for Large-Scale Machine Learning,'' \emph{12th USENIX Symposium on Operating Systems Design and Implementation}, Savannah, GA, USA, 2-4 Nov. 2016.

\end{thebibliography}
\end{document}